\PassOptionsToPackage{table}{xcolor}
\documentclass[sigconf]{acmart}

\setcitestyle{square,sort,comma,numbers}

\usepackage[utf8]{inputenc} %
\usepackage{hyperref}       %
\usepackage{url}            %
\usepackage{booktabs}       %
\usepackage{amsfonts}       %
\usepackage[show]{notes-alt}
\usepackage{nicefrac}       %
\usepackage{microtype}      %
\usepackage{xcolor} %
\usepackage[table]{xcolor}

\usepackage{tcolorbox}
\usepackage{colortbl}

\usepackage{amstext}
\usepackage{amsmath}
\usepackage{enumitem}
\usepackage{algorithm}
\usepackage{algpseudocode}
\usepackage{dblfloatfix}
\usepackage{xspace}
\usepackage{multirow}
\usepackage{wrapfig}
\usepackage{tikz}
\usepackage{pgfplots}
\usepackage{subcaption}
\usepackage{pgfplotstable}
\usetikzlibrary{patterns}
\usepgfplotslibrary{groupplots} 
\usetikzlibrary{matrix, calc} 
\usetikzlibrary{shapes.geometric}

\usetikzlibrary{arrows}           %
\usetikzlibrary{positioning}      %
\usetikzlibrary{fit}              %
\usetikzlibrary{backgrounds}      %

\newcommand{\idnt}{\phantom{ - }}
\newcommand{\sys}[1]{\textsc{Gar}\def\temp{#1}\ifx\temp\empty{}\else\raisebox{-.4ex}{\scriptsize #1}\fi}
\newcommand{\gar}{\textsc{Gar}}

\newcommand{\sgarsys}[1]{\textsc{SlideGar}\def\temp{#1}\ifx\temp\empty{}\else\raisebox{-.4ex}{\scriptsize#1}\fi}

\newcommand{\sysbm}[1]{\sys{}${}_{BM25}$}
\newcommand{\quamsys}[1]{\textsc{Quam}\def\temp{#1}\ifx\temp\empty{}\else\raisebox{-.4ex}{\scriptsize #1}\fi}

\newcommand{\cerberussys}[1]{\textsc{ORE}\def\temp{#1}\ifx\temp\empty{}\else\raisebox{-.4ex}{\scriptsize #1}\fi}

\newcommand{\llama}{RankLLaMA}
\newcommand{\tf}{TFRank}
\newcommand{\qwen}{Qwen}
\newcommand{\mono}{MonoT5}

\newcommand{\rank}{\textsc{Rank1}}

\newcommand{\argmaxm}[1]{%
  \ifthenelse{\isempty{#1}}%
    {\overset{m}{\argmax}}%
    {\underset{#1}{\overset{m}{\argmax}}\, }%
}
\newcommand{\argminm}[1]{%
  \ifthenelse{\isempty{#1}}%
    {\overset{m}{\argmin}}%
    {\underset{#1}{\overset{m}{\argmin}}\, }%
}

\newcommand\mc[1]{\mathcal{#1}}
\newcommand{\sustainable}{{Sustainable Living}\xspace}
\newcommand{\econ}{{Economics}\xspace}
\newcommand{\psychology}{{Psychology}\xspace}
\newcommand{\robotics}{{Robotics}\xspace}
\newcommand{\earth}{{Earth Science}\xspace}
\newcommand{\biology}{{Biology}\xspace}
\newcommand{\stackoverflow}{{Stack Overflow}\xspace}
\newcommand{\pony}{{Pony}\xspace}
\newcommand{\theoremq}{{TheoremQA}\xspace}
\newcommand{\leetcode}{{LeetCode}\xspace}
\newcommand{\aops}{{AoPS}\xspace}

\usepackage{pifont}

\author{Mandeep Rathee}
\orcid{0000-0002-7339-8457}
\affiliation{%
  \institution{L3S Research Center}
  \city{Hannover}
  \country{Germany}  
}
\email{rathee@l3s.de}
\author{Venktesh V}
\orcid{0000-0001-5885-2175}
\affiliation{%
    \institution{Stockholm University}
    \city{Stockholm}
    \country{Sweden}
}
\email{venktesh.viswanathan@dsv.su.se}

\author{Sean MacAvaney}
\orcid{0000-0002-8914-2659}
\affiliation{%
  \institution{University of Glasgow}
  \city{Glasgow}
  \country{United Kingdom}
}
\email{sean.macavaney@glasgow.ac.uk}
\author{Avishek Anand}

\orcid{0000-0002-0163-0739}
\affiliation{%
    \institution{Delft University of Technology (TU~Delft)}
    \city{Delft}
    \country{The Netherlands}
}
\email{avishek.anand@tudelft.nl}

\begin{document}

\title{Reproducing Adaptive Reranking for Reasoning-Intensive IR}




\begin{abstract}

The classical cascading pipeline of retrieve--rerank suffers from a bounded recall problem, stemming from limitations of the first-stage retriever. Most current approaches address the bounded recall problem by improving the first-stage retriever, but this incurs substantial training and inference costs, especially to handle queries that require substantial reasoning. To circumvent the computational costs of reasoning-based retrievers, we replicate the findings of \gar{}, Graph-based Adaptive Reranking, on the BRIGHT reasoning-intensive retrieval benchmark. \gar{} addresses the bounded recall problem by modifying the reranking process itself through iterative exploration of a corpus graph, but it was previously only tested on models designed for topical and question-answering-style queries. Hence, reproduce \gar{} in reasoning-intensive settings with reasoning and non-reasoning reranking models. We observe that the quality of the reranker's signal plays an important role in identifying additional relevant documents within the corpus graph. Overall, we find that \gar{} boosts the effectiveness of reasoning-intensive retrieval across a variety of models while contributing minimally to computational overheads. Ultimately, this work enables more practical deployment of retrieval systems that can address reasoning-intensive queries.

\tikzset{
    docIcon/.pic={
        \filldraw[thick, fill=#1!10, draw=#1!80!black] 
            (-0.2, -0.25) --  %
            (-0.2, 0.25) --   %
            (0.05, 0.25) --   %
            (0.2, 0.1) --     %
            (0.2, -0.25) --   %
            cycle;
        
        \filldraw[thick, fill=white, draw=#1!80!black] 
            (0.05, 0.25) --   %
            (0.05, 0.1) --    %
            (0.2, 0.1) --     %
            cycle;

        \draw[#1!40!black] (-0.15, 0.05) -- (0.0, 0.05);
        \draw[#1!40!black] (-0.15, -0.05) -- (0.15, -0.05);
        \draw[#1!40!black] (-0.15, -0.15) -- (0.05, -0.15);
    }
}

\pgfdeclarelayer{bottom}
\pgfsetlayers{bottom,background,main}

\tikzset{
    dbIcon/.pic={
        \draw[thick, fill=white] (-0.4, -0.6) arc (180:360:0.4 and 0.15) -- (0.4, -0.3) arc (0:180:0.4 and 0.15) -- cycle;
        \draw[thick, fill=white] (-0.4, -0.3) arc (180:360:0.4 and 0.15) -- (0.4, 0.0) arc (0:180:0.4 and 0.15) -- cycle;
        \draw[thick, fill=white] (-0.4, 0.0) arc (180:360:0.4 and 0.15) -- (0.4, 0.3) arc (0:180:0.4 and 0.15) -- cycle;
        
        \draw[thick, fill=gray!20] (0, 0.3) ellipse (0.4 and 0.15);
        
         \draw[thick] (-0.4, -0.3) arc (180:360:0.4 and 0.15);
        \draw[thick] (-0.4, 0.0) arc (180:360:0.4 and 0.15);
        
    }
}

\begin{figure}[] %
    \centering
    \resizebox{\columnwidth}{!}{
    \begin{tikzpicture}[
        font=\sffamily,
        >=stealth, %
        node distance=1cm and 1.5cm,
        block/.style={
            draw, rectangle, rounded corners=3pt, thick,
            minimum width=2.5cm, minimum height=1.5cm, align=center, fill=white
        },
        database/.style={
            cylinder, 
            cylinder uses custom fill, 
            cylinder body fill=white, 
            cylinder end fill=gray!20, 
            shape border rotate=90, 
            aspect=0.25, 
            draw, 
            thick, 
            minimum width=1.5cm, 
            minimum height=1.5cm
        },
        gnode/.style={
            circle, draw, thick, minimum size=0.25cm, inner sep=0pt, fill=white
        },
        doc/.style={
            draw, rectangle, minimum width=2.5cm, minimum height=0.6cm, thick, align=center, font=\small
        },
        docRed/.style={doc, fill=red!10, draw=red!80!black},
        docBlue/.style={doc, fill=blue!10, draw=blue!80!black},
        dashedBlue/.style={->, dashed, color=blue!80!black, thick},
        dashedRed/.style={->, dashed, color=red!80!black, thick},
        standardArrow/.style={->, thick, shorten >=1pt, shorten <=1pt}
    ]
    
        \node (query) {\large Query};
        
        \node [block, right=0.5cm of query, minimum width=2cm, minimum height=1cm, fill=green!10] (retriever) {\large Retriever};
        \draw [standardArrow] (query) -- (retriever);
        
        \coordinate (dbPos) at ($(retriever.south) + (0, -1.5)$);
        \pic (indexPic) at (dbPos) {dbIcon};        
        \node [below=0.8cm of dbPos] {Index};
        
        \draw [standardArrow] (retriever.south) -- ($(dbPos) + (0, 0.45)$);

        \begin{scope}[on background layer]
            \coordinate (ListCenterInt) at ($(retriever.east) + (.8, 0.0)$); 
            \coordinate (docLocInt) at ($(ListCenterInt) + (0.2, 0)$); 
            \draw [standardArrow] (retriever) -- (ListCenterInt);

            \foreach \i in {1,...,6} {
                \pgfmathsetmacro{\ypos}{(3.5 - \i) * 0.8} 
                \coordinate (docLoc\i) at ($(ListCenterInt) + (0.3, \ypos)$); 
                
                \coordinate (docLeft\i) at ($(docLoc\i) + (-0.3, 0)$);
                
                \pic at (docLoc\i) {docIcon={blue}};
                \node[anchor=west,font=\large] (docLabel\i) at ($(docLoc\i)+(0.3, 0.01)$)  {};
                
                \ifnum\i=1 \coordinate (firstDocInt) at (docLoc\i); \fi
                \ifnum\i=6 \coordinate (lastDocInt) at (docLoc\i); \fi
            }
            
            \node [above=0.5cm of firstDocInt, font=\large] {Initial Results};
            
        \end{scope}

        \node [block, right=2cm of docLocInt, minimum width=2cm, minimum height=1cm, fill=orange!10] (reranker) {\large Reranker};
        
        \draw [standardArrow] ($(firstDocInt)!0.5!(lastDocInt) + (0.3, 0)$) -- (reranker.west);
 
        \coordinate [below=2.5cm of reranker] (graphCenter);
        
        \node [gnode, fill=blue!20, draw=blue!80!black] (b1) at ($(graphCenter)+(-1.0, 0.3)$) {};
        \node [gnode, fill=blue!20, draw=blue!80!black] (b2) at ($(graphCenter)+(-0.3, 0.8)$) {};
        \node [gnode, fill=red!20, draw=red!80!black] (r1) at ($(graphCenter)+(0, 0)$) {};
        \node [gnode, fill=red!20, draw=red!80!black] (r2) at ($(graphCenter)+(0.8, -0.15)$) {};
        \node [gnode, fill=red!20, draw=red!80!black] (r3) at ($(graphCenter)+(0.3, -0.8)$) {};
        
        \node [gnode] (w1) at ($(graphCenter)+(-1.3, -0.3)$) {};
        \node [gnode] (w2) at ($(graphCenter)+(-0.8, -0.8)$) {};
        \node [gnode] (w3) at ($(graphCenter)+(1.2, 0.5)$) {};
        \node [gnode] (w4) at ($(graphCenter)+(1.4, -0.3)$) {};
        \node [gnode] (w5) at ($(graphCenter)+(0.65, 1.0)$) {};
        
        \begin{scope}[on background layer]
            \draw [black, thin] (b1) -- (w1);
            \draw [black, thin] (w1) -- (w2);
            \draw [black, thin] (w1) -- (r1);
            \draw [black, thin] (w2) -- (r3);
            \draw [black, thin] (b2) -- (w5);
            \draw [black, thin] (w5) -- (w3);
            \draw [black, thin] (w3) -- (w4);
            \draw [black, thin] (w4) -- (r2);
            \draw [black, thin] (b1) -- (b2);
        \end{scope}
        
        \draw [dashedRed] (b1) -- (r1);
        \draw [dashedRed] (b2) -- (r1);
        \draw [dashedRed] (r1) -- (r2);
        \draw [dashedRed] (r2) -- (r3);

        \node [text=red!80!black, font=\scriptsize, align=left, anchor=west] at ($(r1)+(0.2,1.4)$) {Re-ranker guided \\ traversal};
        \node [text=blue!80!black, font=\scriptsize, align=center] (entryLabel) at ($(b1)+(-0.1, 1.2)$) {\small entry points};
        
        \draw [dashedBlue, bend right=20] (entryLabel.south) to (b1);
        \draw [dashedBlue, bend left=20] (entryLabel.east) to (b2);

        \node [below=4.cm of reranker.south, anchor=south] {Corpus Graph};

        \coordinate (ListCenter) at ($(reranker.east) + (1.5, 0)$); %
        
        \draw [standardArrow] (reranker.east) -- (ListCenter);
        \draw [standardArrow] (reranker.south) -- ++(0, -1.2);

        \foreach \color [count=\i] in {red, blue, red, red, blue, red} {
            \pgfmathsetmacro{\ypos}{(3.5 - \i) * .8}
            
            \coordinate (docLoc) at ($(ListCenter) + (0.5, \ypos)$); %
            
            \pic at (docLoc) {docIcon={\color}};
            
            \ifnum\i=1 \coordinate (topDoc) at (docLoc); \fi
        }
        
        \node [above=0.5cm of topDoc, anchor=south] {\large Final Results};

    \end{tikzpicture}
    }
    \caption{Overview of Graph-based Adaptive Reranking. The blue nodes represent the initial retrieved results, which can also be seen as entry points in the corpus graph. Based on the re-ranker guidance, the corpus graph is traversed. The final results list contains documents from initial results (in blue) as well as new documents from the corpus graph (in red).}
    \label{fig:gar}
\end{figure}
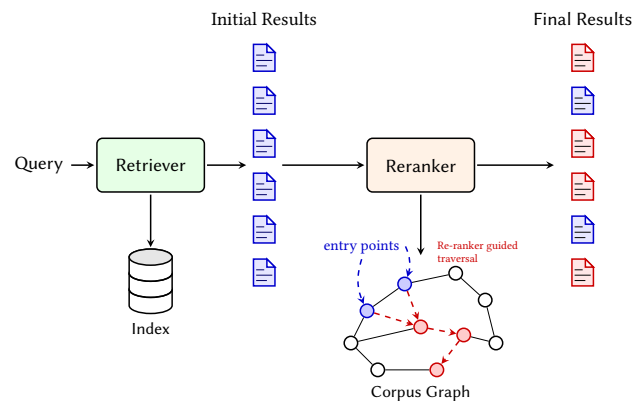

\vspace{0.5em}
\hspace{2.0em}\includegraphics[width=1.25em,height=1.25em]{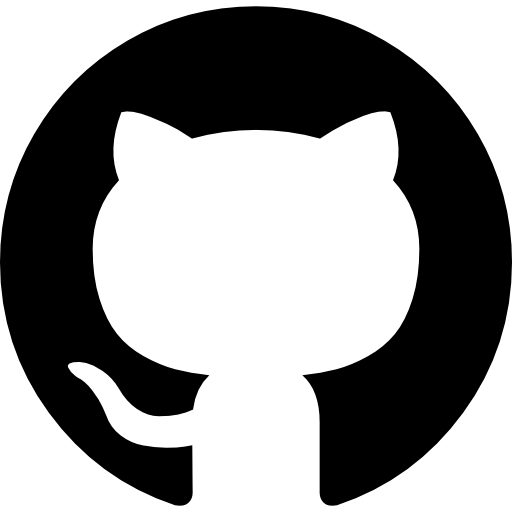}\hspace{.3em}
\parbox[c]{\columnwidth}
{
    \vspace{-.55em}
    \href{https://github.com/Mandeep-Rathee/gar_repro}{\nolinkurl{https://github.com/Mandeep-Rathee/gar_repro}}
}
\vspace{-1.5em}

\end{abstract}

\maketitle

\section{Introduction}
\label{sec:intro}

Ad hoc retrieval has historically focused on matching based on topic relevance~\cite{DBLP:journals/jis/Robertson08}. The increasing capacity of relevance models (especially those powered by Large Language Models (LLMs)~\cite{ma2024fine}) has pushed the field to explore more complicated retrieval tasks, such as those that require \textit{reasoning}\footnote{In this work, we use BRIGHT's~\cite{su2025bright} definition of reasoning-intensive retrieval: search tasks that need more than lexical or semantic similarity to successfully address the provided information need. We acknowledge that this definition has limitations, but leave this discussion for other works.} capabilities (e.g.,~\cite{su2025bright, chenbrowsecomp, killingback2025benchmarking, thakur2025freshstack, weller-etal-2025-followir}). These benchmarking efforts have spawned a variety of new relevance models that aim address this task. Modeling efforts for reasoning-intensive retrieval have focused primarily on cross-encoders as re-ranking models (e.g.,~\cite{weller2025rank1,qwen3embedding,fan2025tfrankthinkfreereasoningenables}) since they enable fine-grained interaction between the query and document. Still, sufficient recall is required from the first-stage retriever to perform well, so there have also been efforts to improve the reasoning capabilities of bi-encoders either by using a reasoning LLM as backbone (without ``thinking'')~\cite{qwen3embedding,fan2025tfrankthinkfreereasoningenables} or using test-time scaling for generating reasoning traces explicitly (``thinking'')~\cite{DBLP:conf/iclr/WellerDLPZH25, zhang-etal-2025-rearank, rankr1, lan2026retro} before generating relevance label. In all cases, however, the trend is clear: larger models (requiring more compute) generally improve the effectiveness of reasoning-intensive retrieval.

Meanwhile, \citet{macavaney2022adaptive} introduced a new perspective on the traditional cascading (retrieve-then-rerank) pipeline. Rather than limiting the re-ranker to only documents identified in the first stage, their approach used a lightweight feedback mechanism from the re-ranker to select which documents to score from a document similarity graph (constructed offline). An overview of this process is shown in Figure~\ref{fig:gar}. This so-called \textit{Graph-based Adaptive Re-ranking} (\gar{}) method consistently improved retrieval effectiveness on traditional benchmarks with minimal computational overhead. The approach has since been adapted to a variety of settings, including improving bi-encoder performance~\cite{kulkarni2023lexically, Kulkarni2024lexboost}, support for listwise re-rankers~\cite{guiding2025rathee, yoon2025listwise}, improved graph construction mechanisms~\cite{rathee2024quam, approximate2025dunn}, alternative prioritization strategies~\cite{ore}, question answering with retrieval augmented generation~\cite{an2026fastinsight,v-etal-2025-sunar}, and a specific adaptation for reasoning-intensive retrieval~\cite{kim2026adaptive}.

The potential benefits of \gar{} are appealing in reasoning-intensive retrieval. If the findings of \gar{} hold in reasoning-intensive retrieval settings, it would buck the trend that more expensive models are needed to improve reasoning capabilities, since effectiveness could be improved with minimal additional cost.
Therefore, in this work, we reproduce\footnote{Technically, \textit{replicate} (using the \href{https://www.acm.org/publications/policies/artifact-review-and-badging-current}{ACM v1.1 definitions}), since we change the experimental setup from testing traditional ad hoc retrieval to reasoning-intensive retrieval.} the findings of \gar{}~\cite{macavaney2022adaptive} to test whether they still hold in reasoning-intensive retrieval settings. We focus on three key findings from the original \gar{} paper: (1) its overall effectiveness, (2) its robustness across rerankers, and (3) its robustness across its hyperparameters. We confirm all three: (1) \gar{} is highly-effective in reasoning-intensive retrieval, (2) it works across a variety of reasoning rerankers (Qwen~\cite{qwen3embedding}, TFRank~\cite{fan2025tfrankthinkfreereasoningenables}, and \rank~\cite{weller2025rank1}), and (3) it is largely robust to its batch size and number of neighbor hyperparameters in the reasoning-intensive settings. Beyond these reproduction findings, we observe that \gar{} can overcome model size trends (e.g., TFRank-4B with \gar{} performs better than TFRank-8B without \gar{} by average nDCG@10 performance on BRIGHT). We also observe that the strongest models usually benefit the most from \gar{} (e.g., Qwen-4B benefits \textit{more} with \gar{} than Qwen-0.6B). This result is important because it shows that findings are \textit{composable}: improvements in relevance modeling stack upon (and actually accelerate) those from \gar{}.

\definecolor{paleYellow}{RGB}{255, 250, 235}
\definecolor{paleGreen}{RGB}{235, 248, 235}
\definecolor{textGreen}{RGB}{34, 139, 34}
\definecolor{tagRed}{RGB}{230, 100, 100}
\definecolor{tagGreen}{RGB}{50, 205, 50}
\definecolor{markerRed}{RGB}{233, 150, 149}
\definecolor{markerGreen}{RGB}{0, 180, 0}
\definecolor{grayText}{RGB}{100, 100, 100}

\begin{figure}[t]
    \centering
    \begin{tcolorbox}[
        colback=paleYellow, 
        colframe=orange!50!yellow, 
        boxrule=0pt, 
        leftrule=2pt, 
        sharp corners,
        arc=0pt,
        width=\columnwidth,
        boxsep=2pt,
        left=2pt, right=2pt, top=2pt, bottom=2pt
    ]
        \textbf{Query}
        \par\vspace{2pt}
        \small
        Claim in article about why insects are attracted to light... Heat radiation as an attractive component is refuted by the effect of LED lighting... Could they for example be evolutionarily programmed to associate light with heat? So that even though they don't encounter heat near/on the LEDs they still "expect" to?
    \end{tcolorbox}

    \vspace{2pt}

    \begin{tcolorbox}[
        colback=paleGreen, 
        colframe=markerGreen, 
        boxrule=0pt, 
        leftrule=2pt, 
        sharp corners, 
        arc=0pt,
        width=\columnwidth,
        boxsep=2pt,
        left=2pt, right=2pt, top=2pt, bottom=2pt
    ]
        \setlength{\tabcolsep}{1pt}
        \begin{tabular*}{\linewidth}{@{\extracolsep{\fill}} l l l l}
            {\small \color{grayText} Initial Recall} & {\small \color{grayText} Neighborhood Recall} & {\small \color{grayText} Total Recall} & {\small \color{grayText} \#New documents} \\
            \textbf{\scriptsize \color{textGreen} 0.0} & \textbf{\scriptsize \color{textGreen} 100.0} & \textbf{\scriptsize \color{textGreen} 100.0} & \textbf{\scriptsize \color{textGreen} 5} \\
        \end{tabular*}

     \end{tcolorbox}

    \vspace{4pt}

    \centering    
    \begin{tikzpicture}
        \begin{axis}[
           axis background/.style={fill=gray!10},
            width=\columnwidth,
            height=4.5cm,
            xmin=0, xmax=130,
            ymin=0, ymax=50,
            y dir=reverse,
            axis line style={draw=none},
            tick style={draw=gray!30, major tick length=3pt},
            xtick={0, 20, 40, 60, 80, 100, 120},
            xticklabel style={font=\small, color=grayText},
            xlabel={\small Neighbours},
            xlabel style={yshift=0.1cm, color=grayText},
            ytick={0, 25},
            ytick={0,10,20,30,40,50},
            yticklabel style={font=\small, color=grayText},
            ylabel={\small Initial Retrieved Documents},
            ylabel style={color=grayText},
            xtick pos=left, 
        ]
            \addplot[
                scatter, 
                only marks, 
                mark=star, 
                mark size=2pt,
                mark options={fill=markerGreen, color=markerGreen, scale=1}
            ] coordinates {(1, 22) (4, 22) (32, 22) (52, 39) (89, 39)};

            \addplot[
                scatter, 
                only marks, 
                mark=square*, 
                mark size=2pt,
                mark options={fill=tagRed, scale=1.0}
            ] coordinates {
                (85, 14) (109, 28) (109, 29)
            };
        \end{axis}
    \end{tikzpicture}

\vspace{2pt}
\begin{center}
    \resizebox{0.85\columnwidth}{!}{%
        \begin{tikzpicture}
            \draw[color=black, fill=gray] (0,0) plot[mark=star, mark size=3pt] coordinates {(0,0)};
            \node[right, font=\small, text=grayText] at (0.2, 0) {Relevant Document};

            \draw[color=black, fill=gray] (4.0,0) plot[mark=square*, mark size=2.5pt] coordinates {(4.0,0)};
            \node[right, font=\small, text=grayText] at (4.2, 0) {Duplicate Relevant Document};
        \end{tikzpicture}%
    }
    \end{center}
    \caption{Qualitative example for showing the position of relevant documents in the neighborhood for a query from the Biology subset of BRIGHT. The initial retrieval BM25 fails to find any relevant document (hence Recall is 0); however, the 1-hop neighbors contain all relevant documents (5 new and 3 duplicates) and demonstrate Recall of 100\%. }
    \label{fig:query_analysis}
\end{figure}
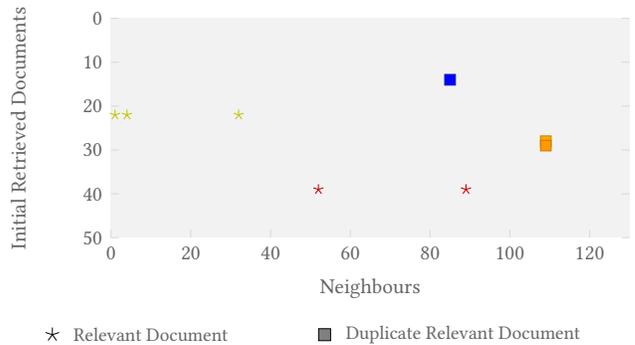

Overall, our reproduction study findings have important implications on future directions in reasoning-intensive retrieval. We show that re-ranking ``smarter'' using \gar{} has cost advantages over ``bigger'' re-rankers. For instance, in our results, we observe that \tf-1.7B augmented with \gar{} achieves better performance than \qwen{}-4B (e.g: upto \textbf{27.72\%} gains in nDCG@10 on biology subset) and \rank{}-7B (e.g: upto \textbf{26.51\%} gains in nDCG@10 on earth sciences subset). Furthermore, the \gar{} and larger re-ranker strategies are composable.

\section{Background and Preliminaries}
\label{sec:background}

Complex transformer based ranking models and, more recently instruction tuned Large Language Models (LLMs) have advanced document ranking. LLM-based rankers are highly effective in providing precise relevance estimates, but computationally expensive. More recently, reasoning-based LLMs have been adopted for reasoning-intensive IR tasks, which are further expensive and lead to high latency during inference. We contextualize our work into non-reasoning (traditional) and reasoning-based retrieval and reranking methods.

\subsection{Traditional Retrieve-then-Rerank and Adaptive Retrieval}

Traditional IR pipelines usually adopt a telescoping setting, which entails a fast but imprecise estimation of relevance using first-stage retrieval, followed by heuristic filtering of top-$k$ documents which are then sent to a reranker for a slow but precise estimation of relevance.

Traditional lexical retrieval methods, such as BM25~\cite{bm25,splade, wang2011cascade}, are efficient for first-stage retrieval, but the resulting relevance scores are imprecise. 
In modern retrieval systems, dense retrievers~\cite{lin2021batch, karpukhin-etal-2020-dense}, learned sparse~\cite{DBLP:conf/sigir/MacAvaneyN0TGF20b,splade}, and hybrid sparse-dense ensembles are used for first-stage retrieval~\cite{ cormack2009reciprocal,bruch2023analysis, chen2022out,wang2021bert}. Meanwhile, complex re-rankers such as cross-encoders, which employ transformers, provide better relevance estimates. However, due to their computational requirements and overhead in latency, they are leveraged in the final stage of the telescoping setting \cite{ore} to rank a limited set of top-$k$ documents filtered based on retrieval scores. However, as observed in prior works \cite{rathee2024quam}, this leads to a ``bounded recall" problem, where documents that are ignored based on imprecise estimation of relevance using first-stage retrieval scores are never considered for re-ranking. To recover such relevant documents, methods like \gar{} \cite{macavaney2022adaptive} were proposed, which adaptively retrieve documents guided by a corpus graph. \gar{} builds upon the \textit{clustering hypothesis} \cite{jardine1971use}, which states that documents that are close to each other help address the same query. \gar{} constructs a neighborhood graph comprising of top-$k$ neighbors for each document and employs a cross-encoder to rank neighbors of already ranked documents adaptively. Kulkarni et. al. \cite{kulkarni2023lexically} explore employing a bi-encoder for efficiency instead of a cross-encoder for adaptive ranking. There are other types of further fine-grained graphs, e.g., knowledge graphs,~\cite{edge2024local} creation methods. In our study, following the original paper \gar{}, we stick to corpus graphs where nodes are the documents.

With advances in Large Language Models (LLMs) on a wide range of language tasks, they were also extended to document ranking tasks to obtain better relevance estimates. Several pointwise rankers like RankLLama~\cite{ma2024fine} and listwise rankers like RankZephyr~\cite{pradeep2023rankzephyr}, RankGPT~\cite{sun2023chatgpt} have been proposed.

However, reasoning-intensive tasks like BRIGHT \cite{su2025bright} require complex query and relevance understanding. Hence, they require intensive Chain-of-Thought~\cite{wei2022chain} based reasoning to accurately model the semantics of the query and the documents to arrive at the right relevance estimate. We detail on some of the approaches for reasoning-intensive rankers in Section~\ref{sec:rel_work_reasoning}.

\subsection{Reasoning-based Retrieval and Reranking}
\label{sec:rel_work_reasoning}
Apart from traditional retrieval tasks, more recently, much interest has been garnered around reasoning-intensive tasks \cite{thakur2025freshstack,su2025bright}. Rather than simple determination of similarity through semantic relatedness and surface form matching, these tasks require deliberate reasoning. For instance, a coding task may require the ranker to match based on the underlying logic required to solve the question, rather than just keyword and semantic matching. Hence, such tasks require retrievers and rankers that are capable of deeper reasoning to find accurate matches.

Hence, in recent years, with advances in LLMs and instruction tuning, retrieval systems have advanced beyond simple phrase-level or surface form matching \cite{weller2025rank1,weller-etal-2025-followir,asai-etal-2023-task,muennighoff2025generative}. These models, like TART and FollowIR, usually
use instruction-based data in their training data, so that they learn to adapt to new user instructions.  These models understand the meaning/intent underlying the user query better than just phrase-level matching and are also capable of following instructions.
Advances in decoder LLMs led to their increased adoption for document ranking tasks through Low Rank Adaptation (LoRA) for efficient fine-tuning. Hence, LLM based rankers have been adopted for pointwise \cite{ma2024fine} (\llama{}), pairwise \cite{sinhababu-etal-2024-shot}, and listwise settings \cite{pradeep2023rankzephyr} (RankZephyr). While they demonstrate better query understanding and strong performance compared to existing rankers, they still fall short in reasoning-intensive tasks. Hence, more recently, reasoning-based LLMs have been adapted for retrieval and ranking tasks that require deeper reasoning. 

For instance, \rank~\cite{weller2025rank1} introduces test-time compute in IR for allotting more compute for enhancing re-ranking of documents. \rank~\cite{weller2025rank1} first obtains 635,000 reasoning traces from a large reasoning model like Deepseek-R1~\cite{guo2025deepseek} on MS-MARCO. These reasoning traces are then used for distillation to smaller reranker LLMs. Thinking free ranker TFRank \cite{fan2025tfrankthinkfreereasoningenables} further enhances the training paradigm by proposing an efficient reasoning-based ranker leveraging reasoning models. Their training recipe entails a multi-task learning strategy including fine-grained relevance scores for pointwise, listwise, and pairwise paradigms distilled from a Deepseek-R1 reasoning model. REARank \cite{zhang-etal-2025-rearank} extends reasoning language models to listwise ranking, with explicit reasoning that is incentivized using reinforcement learning, under data-scarce scenarios with a data augmentation strategy, outperforming larger scale LLMs on reasoning-intensive document ranking tasks. Rank-R1 \cite{rankr1} trains an LLM as a setwise reranker using RL.
It simplifies the task of finding the most relevant passage index, relying on a sparse matching-based binary reward signal. While these models have significantly improved ranking capabilities, they are still limited by the ``bounded recall" problem. Hence, in this work, we analyze the effect of reproducing \gar{} in a reasoning-intensive setup.

\subsection{A Recap of \gar{}}
Figure~\ref{fig:gar} shows the \gar{} visual representation. \gar{} solves \textit{bounded-recall} problem by adaptively reranking documents. For a given reranking budget $c$, it starts with selecting $b$ documents from the initially retrieved documents list, denoted by $R_{0}$. These first $b$ documents can be seen as entry points in the corpus graph, as shown in Figure~\ref{fig:gar}. Then, a pointwise reranker (a cross encoder such as MonoT5) is applied to rerank these $b$ documents, and then these documents are added to the final rank list (denoted as $R_1$). Then, a graph frontier ($F$) is prepared using the neighbors of already reranked documents. For this, the \textit{offline} built corpus graph is traversed. In the graph frontier, each neighbor is assigned the score of the source document given by the ranker. By this, the neighbors of highly reranked documents are prioritized based on the \textit{clustering hypothesis}, and therefore the feedback from the reranker (quantified by relevance score) becomes crucial. 
For the next batch, the top $b$ documents from the graph frontier $F$ are selected and ranked. Next, these documents are added to the final rank list $R_1$. Then, the graph frontier $F$ is updated with the neighbors of these newly ranked documents. \gar{} keeps selecting batches from $R_0$ and graph frontier $F$ alternatively until the reranking criteria is met, i.e, $|R_1|=c$. 
The final rank list $R_1$ contains documents from the initial results and the graph (as shown in Figure~\ref{fig:gar}).
\gar{} shows high effectiveness on MSMARCO-passage collection and test sets such as TREC DL19 and 20. However, these test sets consist of web queries that are short and ambiguous. In our work, we reproduce \gar{} on the BRIGHT benchmark, where queries are verbose, and relevance between a query and a document is defined by reasoning between them.

\section{Experimental Setup}
While reproducing \gar{}, we address the following research questions through extensive experiments:

\textbf{RQ1}: Is \gar{}'s effectiveness reproducible in reasoning-intensive information retrieval task?  

\textbf{RQ2}: What is the effect of feedback signals from different rerankers in \gar{}? 

\textbf{RQ3}: Does \gar{} remain robust to selection of hyperparameters?

\begin{table}[!t]
\caption{BRIGHT benchmark statistics.}
\centering
\begin{tabular}{l|rrr|rr}
\toprule
& \multicolumn{3}{c|}{Total Number} & \multicolumn{2}{c}{Avg. Length}  \\
\cmidrule{2-6}
\bf{Dataset} & \multicolumn{1}{c}{$\mathbf{Q}$} & \multicolumn{1}{c}{$\boldsymbol{\mathcal{D}}$} & \multicolumn{1}{r}{$\boldsymbol{\mathcal{D}^+}$} & \multicolumn{1}{c}{$\mathbf{Q}$} & \multicolumn{1}{r}{$\boldsymbol{\mc{D}}$}  \\
\midrule
 \multicolumn{6}{c}{\textit{StackExchange}} \\
\midrule
\biology & 103 & 57,359 & 3.6 & 115.2 & 83.6  \\
\earth & 116 & 121,249 & 5.3 & 109.5 & 132.6  \\
\econ & 103 & 50,220 & 8.0 & 181.5 & 120.2 \\
\psychology & 101 & 52,835 & 7.3 & 149.6 & 118.2  \\
\robotics & 101 & 61,961 & 5.5 & 818.9 & 121.0  \\
\stackoverflow & 117 & 107,081 & 7.0 & 478.3 & 704.7  \\
\sustainable & 108 & 60,792 & 5.6 & 148.5 & 107.9  \\
\midrule
\multicolumn{6}{c}{\textit{{Coding}}} \\
\midrule
\leetcode & 142 & 413,932 & 1.8 & 497.5 & 482.6  \\
\pony & 112 & 7,894 & 22.5 & 102.6 & 98.3  \\
\midrule
\multicolumn{6}{c}{\textit{{Theorems}}} \\
\midrule
\aops & 111 & 188,002 & 4.7 & 117.1 & 250.5   \\
\theoremq-Q & 194 & 188,002 & 3.2 & 93.4 & 250.5   \\
\theoremq-T & 76 & 23,839 & 2.0 & 91.7 & 354.8  \\
\bottomrule
\end{tabular}
\label{tab:statistics}
\end{table}

\subsection{Datasets}
We use the recently proposed reasoning-intensive retrieval benchmark, BRIGHT~\cite{su2025bright}, for our evaluation. The benchmark consists of multiple subsets or subtaks from domains such as Stack Exchange posts, Coding, and Mathematical Theorems. For example, in Stack Exchange, each query is a post from the user, and accepted answers or webpages whose link is provided in the answer are relevant documents. Table~\ref{tab:statistics} provides the statistics on these subsets or subtasks. 
We build a sparse PISA~\cite{DBLP:conf/sigir/MalliaSMS19} index for each subset of the BRIGHT. To build the corpus graph, we use the sparse index and encode lexical similarities between documents, such as the BM25 score. Similarly to the original paper \gar{}, we treat the document as a query and add \textit{top-k} retrieved results (given by BM25) as neighbors in the graph. This corpus graph is built \textit{offline}, and the cost of building such graphs is described in the original paper. The original paper \gar{} built graphs using lexical (BM25) and semantic (TCT-ColBERT~\cite{lin2021batch}) similarities. Due to limited resources, we reproduce \gar{} only with the BM25-based corpus graph, given its robust performance on BRIGHT. We leave other types of corpus graph formulation methods, such as semantic similarity based using dense encoders, for future work.
We release all details on building indexes and corpus graphs along with our code repository. 

\begin{table*}[t!]
\caption{Ranking performance in nDCG@10 of different type of rerankers. All models rerank the top-$100$ retrieved documents from BM25 using the original query. w/ \gar{} represents the performance of the ranker when reranking is done using \gar{}. Results in bold represent where the \gar{} variant outperforms the standard reranking pipeline. }
\centering
\resizebox{\textwidth}{!}{
\begin{tabular}{l|rrrrrrr|rr|rrr|r}
\toprule
& \multicolumn{7}{c|}{StackExchange} & \multicolumn{2}{c|}{Coding} & \multicolumn{3}{c|}{Theorem-based} & \multirow{2}{*}{\centering Avg.}\\
\cmidrule(r){2-8} \cmidrule(r){9-10} \cmidrule(r){11-13}
& Bio. & Earth. & Econ. & Psy. & Rob. & Stack. & Sus. & Leet. & Pony & AoPS & TheoQ. & TheoT. \\
\midrule
BM25  & 17.2  & 25.6  & 14.0 &12.7 & 14.9 &17.4 & 12.6 & 13.2 & 8.8 & 3.0 & 6.7 & 4.9 & 12.6 \\
\midrule
\multicolumn{14}{c}{\textit{Non-Reasoning Rerankers}} \\
\midrule
MonoT5-base & 9.9 & 16.6  & 7.9  &7.4  & 4.5 & 6.2  & 10.2& 6.9 &  7.0&  2.9& 8.0 & 3.3 &7.6 \\ 
\idnt w/ \gar & 9.2 & 16.3 & \textbf{8.0}  & \textbf{7.8} & 4.4 & 5.9  &9.7 & \textbf{7.0} & 6.6 &2.4  &\textbf{9.8} &\textbf{4.1} & 7.6 \\
MonoT5-3B & 9.4   & 17.2 & 10.3 & 7.4 & 7.5& 8.8 & 10.6 & 7.7 &  6.3& 2.7 &  7.8& 4.9 & 8.4\\
\idnt w/ \gar & \textbf{9.5} & 16.7 & \textbf{11.3}  & \textbf{7.5} & \textbf{8.9} & 8.8  &10.5 & 7.6 & 3.9 & \textbf{2.9} &\textbf{9.5} &\textbf{5.2} & \textbf{8.5} \\
\cmidrule{2-14}
RankLLaMA-7B & 17.4  & 29.0  & 9.9 & 14.2 &15.7 & 7.2 & 16.6 &8.7  &  23.5 & 1.6 &  4.7& 3.6 & 12.7 \\
\idnt w/ \gar & 10.0 & 17.5 & 8.1  & 9.1 & 15.5 & 6.2 &\textbf{16.8} &6.5  & \textbf{29.8} &1.6  &3.2 &3.0 & 10.6  \\
\midrule
\multicolumn{14}{c}{\textit{Reasoning but Non-Thinking Rerankers}} \\
\midrule
 Qwen-0.6B & 19.8  &25.2  & 14.2 & 18.7& 17.1& 19.2 & 16.2 &  17.1 &  4.9 &  2.5&  9.3&  12.2&14.7 \\
\idnt w/ \gar & \textbf{20.8} & \textbf{25.4} & \textbf{15.2}  & 17.5 & 16.4 & 18.4  & \textbf{16.8} & 16.8 & \textbf{6.3} & \textbf{2.9} &\textbf{10.9} &\textbf{16.2} &\textbf{15.3}  \\
Qwen-4B & 25.7  & 36.4 & 19.2 & 24.4 & 26.7& 28.4 & 24.1 & 17.9 & 19.9 & 3.3 & 11.4 & 12.4 & 20.8\\
\idnt w/ \gar & \textbf{28.5} & \textbf{37.6} & \textbf{22.3}  & \textbf{25.9} & \textbf{27.4} & 28.1  &\textbf{26.4} & \textbf{18.3} & \textbf{20.0} &\textbf{4.0}  & \textbf{16.0} &\textbf{19.0} & \textbf{22.8} \\
\qwen-8B &  28.5 &33.6 &20.7 &25.8 & 29.2 & 26.7 &23.6  & 16.1 & 18.7 & 3.6 & 11.2 &12.2  & 20.8\\
\idnt w/ GAR & \textbf{34.3} & \textbf{34.3} &\textbf{24.5}&\textbf{28.8} & \textbf{30.7} & 25.7 & \textbf{25.7} & \textbf{16.9}  & 16.3 & 3.6 & \textbf{15.0} & \textbf{17.8} & \textbf{22.8} \\
\cmidrule{2-14}
TFRank-0.6B & 24.1 & 32.9 &18.4 & 22.1& 15.4& 20.2 & 19.4 &  17.6& 6.6 &1.8 &10.0  & 8.9 & 16.5\\
 \idnt w/ GAR &\textbf{27.5} &\textbf{34.2} &\textbf{19.6} & \textbf{26.7} &\textbf{16.3}  & 19.4 & 21.2 & 17.3 & \textbf{8.9} & \textbf{2.0} & \textbf{11.5} & \textbf{11.1} &\textbf{18.0} \\
TFRank-1.7B& 32.3  & 41.2 & 18.1 &27.0 & 17.2 & 25.8 &21.2  & 14.1 & 16.4 & 1.5 & 9.4 & 12.0 &19.7 \\
 \idnt w/ GAR &\textbf{36.4} & \textbf{43.9}  & \textbf{21.6} & \textbf{28.4} & 16.9 & 24.6 & \textbf{23.3} & \textbf{14.2} & \textbf{21.6} & \textbf{1.7} & \textbf{11.0} & \textbf{17.1} & \textbf{21.7}\\
TFRank-4B & 36.4  & 44.2  & 21.1 & 29.5& 22.7 & 27.2 & 27.8 &  19.0&  25.3& 1.4 & 9.1 & 12.5  & 23.0\\
 \idnt w/ GAR &\textbf{45.1} & \textbf{46.8} & \textbf{23.9} & \textbf{32.6} & \textbf{22.7} & 26.0  & \textbf{32.1} & \textbf{20.1} & 24.5 & 1.5 & \textbf{11.2} & \textbf{18.2} &\textbf{25.4} \\
TFRank-8B &  35.8 & 45.8  & 22.7 & 27.2& 23.7& 27.9 & 30.2 & 19.0 & 24.8 & 1.5 &  12.7& 12.7 & 23.7 \\
 \idnt w/ GAR &\textbf{44.8} &\textbf{48.9}  &\textbf{25.3}  & \textbf{28.0} & \textbf{26.5} &27.6  & \textbf{32.9} & \textbf{20.3} & 24.4 & \textbf{1.5} & \textbf{19.0} &\textbf{19.0} & \textbf{26.5}\\

\midrule
\multicolumn{14}{c}{\textit{Reasoning and Thinking Rerankers}} \\
\midrule
\rank-0.5B & 11.6  & 13.6 & 5.3 & 11.9 & 7.1 & 9.6 & 12.0 & 2.7 & 12.2 & 1.3 & 4.0 & 2.5 & 7.8 \\
\idnt w/ GAR & \textbf{13.4} & 9.8 & \textbf{5.3} & 8.7& 6.0 &8.4  & 10.3 &2.2  & 11.6 & \textbf{1.6} & \textbf{4.3} & 1.9 & 7.0\\
\rank-1.5B & 20.9  & 18.4 & 8.5 & 18.1 & 11.0 &10.2 &14.0 & 5.6 & 11.5 & 1.8 & 5.8 &  6.4 & 11.0\\
\idnt w/ GAR & \textbf{24.0}  & 15.6 & 8.3 & 16.7 &8.7  & 8.7  & 11.3 & 3.9 & 9.9 & 1.6 & \textbf{6.4} & \textbf{7.7} & 10.2 \\
\rank-7B & 34.0  & 33.8 & 17.9 & 25.6 & 16.7& 20.3 & 24.0 & 9.5 & 25.9 & 4.6 &  10.5&  13.2& 19.7 \\
 \idnt w/ GAR & \textbf{40.2} & \textbf{34.7}  & \textbf{20.1} & \textbf{28.4} & \textbf{17.7} & 18.9   & \textbf{25.0}  & 8.3  & 24.4  & 4.0  & \textbf{12.5} & \textbf{17.5}  & \textbf{21.0} \\

\bottomrule
\end{tabular}
}
\label{tab:bright}
\end{table*}

\subsection{Retrieval and Rerankers} 
We use BM25~\cite{bm25} as initial retrieval (with $k_1$=0.9, $b$=0.4)\footnote{We follow the BRIGHT implementation for these parameters.} for each subset, and retrieve the top $c$ documents exhaustively. For reranking, we choose a wide range of methods, including reasoning and non-reasoning models. We use the definition of reasoning in the Introduction section for the classification of these models. In non-reasoning, we use MonoT5~\cite{nogueira2020document} and RankLLaMA~\cite{ma2024fine} models. Further, we divide the reasoning models into two categories. First, those that do not ``think'' at test-time (e.g., no reasoning traces are generated before giving the relevance label), such as TFRank~\cite{fan2025tfrankthinkfreereasoningenables} (\texttt{/no think} variants) and Qwen3~\cite{qwen3embedding} family rerankers. These models also follow instructions given the prompt; however, do not generate reasoning traces explicitly. For \qwen, we follow the prompts~\footnote{\url{https://github.com/augustinLib/SPIKE/tree/main/data/embedding_instruction/gte-Qwen1.5-7B-instruct}} used in previous works~\cite{lee2025imagine,su2025bright} and models released by Qwen~\cite{qwen3embedding}. We use \qwen-0.6B, 4B, and 8B variants. For \tf, we use fusion of binary judgments (\textit{yes/no}) and fine-grained score (0-4) given by a pointwise ranking setting. We use the models based on GRPO fine-tuning on Qwen3 models and the \textit{huggingface} weights released by the authors of \tf. We use \tf-0.6B, 1.7B, 4B, and 8B variants. Second, those reasoning models that ``think'' at test time and generate reasoning traces before generating relevance labels (\texttt{true/false}). In this category, we use \rank~\cite{weller2025rank1} family rerankers. \rank{} also follows the instruction in the prompt, where the definition of relevancy can be provided to the reasoning model. We follow the prompt for each subtask from the official repository of \rank~\footnote{https://github.com/orionw/rank1/blob/main/prompts.py} and use \rank-0.5B, 1.7B, and 7B variants. Although the authors of \rank{} released 14B and 32B variants, we discard those because of limited resources and their diminishing returns on BRIGHT.  

We select these rerankers on the basis of their effectiveness on web-style TREC test sets and reasoning-intensive benchmarks. We believe that these rerankers represent most of the existing rerankers.

\subsection{Baselines}
We compare \gar{} with the BM25 retriever, which is a strong baseline on the BRIGHT benchmark. We also compare the performance of \gar{} with the standard reranking pipeline, where BM25 initially retrieves $c$ documents and then a neural ranker (pointwise) reranks. We measure the ranking performance by nDCG@10 and retrieval performance by Recall@c, where $c$ is the retrieval depth or re-ranking budget. We also report the average performance on the BRIGHT benchmark, in line with other works.

\subsection{Other Hyperparameters}
We select the retrieval depth or reranking budget $c$ to $50$ and $100$. As reported in the original paper, \gar{} is robust to the selection of hyperparameters used in the algorithm, mainly the batch size $b$ and number of neighbors $k$ in the graph. For our main reproducibility experiments, we set $b=16$ and $k=16$. We also do ablation on the variation of $b$ and $k$ in Section~\ref{sec:ablation}. For ablation experiments, we set the ranking budget $c=50$ and only $3$ subtasks of BRIGHT due to limited resources and computationally heavy rerankers.

\section{Results and Analysis}

\subsection{Effectiveness of \gar{} on BRIGHT}
\label{sec:results_rq1}
\label{sec:effectiveness}
To answer \textbf{RQ1} and \textbf{RQ2} (partially), we compare different reasoning-based and non-reasoning rankers in settings augmented with \gar{} and without \gar{}. The results of ranking performance are shown in Table \ref{tab:bright} as indicated by nDCG@10, and the retrieval performance is shown in Figure \ref{fig:recall@100} as indicated by Recall@100. We also present a detailed analysis of results for answering \textbf{RQ2} by comparing the performance of \gar{} when using LLM rankers of different parameter scales in Section~\ref{sec:model_comparison}.

\definecolor{garblue}{HTML}{4C72B0}
\definecolor{bm25orange}{HTML}{C44E52}
\definecolor{reasonirgreen}{HTML}{55A868}
\definecolor{qwencoral}{HTML}{C19A6B}
\definecolor{bm25reasonmutedGold}{HTML}{C19A6B}
\definecolor{tfrank}{HTML}{C19A6B}

\definecolor{deepTeal}{HTML}{39737C}
\definecolor{mutedPurple}{HTML}{8172B3}

\begin{figure*}
\centering
\begin{tikzpicture}
\begin{axis}[
    width=16cm,
    height=7cm,
    ybar=2pt, %
    bar width=8pt,
    title={},
    ylabel={Recall@100},
    xlabel={Subset},
    symbolic x coords={Avg., Bio., Earth., Econ., Psy., Rob., Sus., Stack.},
    xtick=data,
    ymin=0, ymax=80, %
    ytick={10, 20, 30, 40, 50, 60, 70},
    enlarge x limits=0.1,
    legend columns=2,
    legend style={at={(0.98,0.98)}, anchor=north east, nodes={scale=0.8, transform shape}},
    nodes near coords,
    every node near coord/.append style={font=\tiny, /pgf/number format/fixed, /pgf/number format/precision=1},
    legend cell align={left},
    axis line style={black},
    major tick style={black},
    tick label style={font=\small},
    label style={font=\large}
]

\addplot[fill=bm25orange, draw=none] coordinates {
    (Avg., 45.9) (Bio., 46.8) (Earth., 58.4 ) (Econ., 40.2) 
    (Psy., 40.2) (Rob., 45.3 ) (Sus., 44.2) (Stack., 46.2)
};

\addplot[fill=qwencoral, draw=none] coordinates {
    (Avg., 53.7) (Bio., 59.4) (Earth., 63.4) (Econ., 50.0) 
    (Psy., 45.3) (Rob., 51.0) (Stack., 52.3) (Sus., 54.3)
};

\addplot[fill=reasonirgreen, draw=none] coordinates {
    (Avg., 52.3) (Bio., 64.0) (Earth., 64.4) (Econ., 45.6) 
    (Psy., 44.5) (Rob., 46.8) (Sus., 50.2) (Stack., 50.9)
};

\addplot[fill=garblue, draw=none] coordinates {
    (Avg., 55.9) (Bio., 64.4) (Earth., 68.0) (Econ., 50.6) 
    (Psy., 49.6 ) (Rob., 49.6) (Sus., 54.8) (Stack., 54.4)
};

\legend{ BM25, Qwen-0.6B w/ \gar{}, Rank1-1.5B w/ \gar{}, TFRank-1.7B w/ \gar{}}

\end{axis}
\end{tikzpicture}
\caption{Retrieval performance in Recall@100 across different methods on the StackExchange subset of BRIGHT. First, BM25 retrieves the top-$100$ documents using the original query, then the adaptive reranking (\gar{}) is done using different rankers. }
\label{fig:recall@100}
\end{figure*}
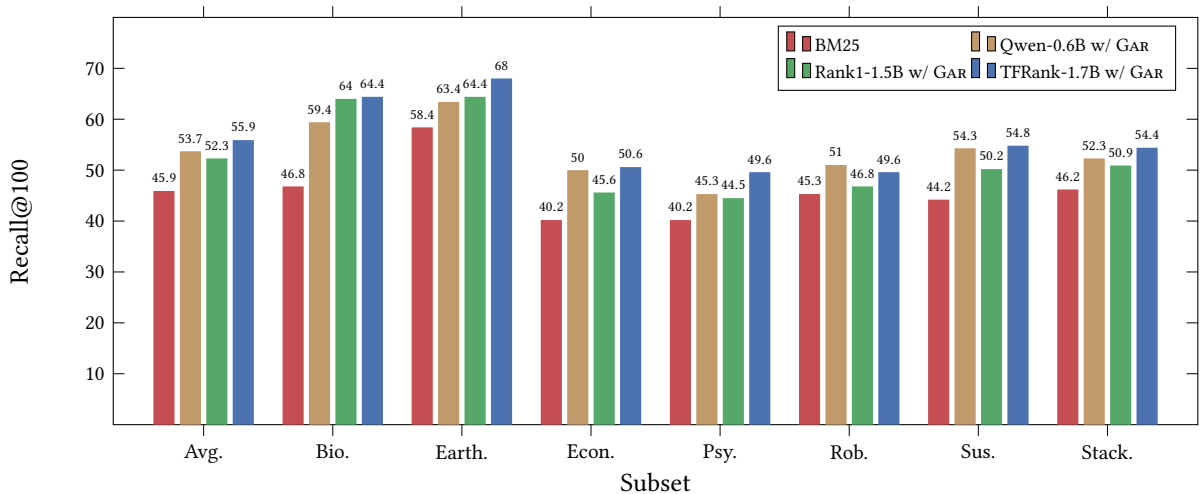

\subsubsection{Retrieval and Ranking performance of \gar{} with stronger rankers}
From Table \ref{tab:bright}, we observe that both medium and large scale (in terms of parameters) reasoning-based but non-thinking rerankers like \qwen{}- (4B, 8B) series, \tf{} series, and reasoning-based thinking rerankers like \rank{} (7B), when augmented with \gar{}, observe improvements in ranking performance. For instance, on the Biology collection, we observe upto \textbf{25.1\%} gains in nDCG@10 when \tf-8B is augmented with \gar{} compared to its ranking performance without \gar{}. We also observe upto \textbf{49.6\%} gains on TheoremQ and TheoremT collections. This is primarily because the initial BM25 retrieved results only capture a limited number of relevant documents, limiting the ranker due to the ``bounded recall" problem \cite{rathee2024quam,ore}. However, \gar{} aids in finding more relevant documents from the neighborhood in the corpus graph with the help of relevance feedback signals from the ranker, circumventing the bounded recall issue. This is also evident by analyzing the retrieval performance as measured by Recall@100 in Figure \ref{fig:recall@100}. Due to limited space, the plots only include retrieval performance analysis from the rankers \tf-1.7B, \qwen{}-0.6B, and \rank{}-1.5B, though we observe similar gains for other strong rankers. We observe that when augmented with \gar{} Recall@100 improves upto \textbf{36.8\%} (performance of \rank{}-1.5B augmented with \gar{}, on Biology collection). We observe considerable gains in Recall@100 when employing \gar{} with feedback from different LLM rankers (\tf, \qwen{}) as observed in Figure \ref{fig:recall@100} due to its ability to capture relevant documents from the neighborhood. These additional relevant documents are then ranked higher by the ranker, resulting in higher nDCG@10 as observed in our results. Note that while \gar{} helps in capturing more relevant documents, it is the strength of the ranker that determines if these documents are surfaced higher up in the results.  

The gains are observed across models that perform thinking by generating a detailed justification for the relevance score during inference, like \rank{}, and also models that do not perform any thinking. These results demonstrate that \gar{} serves as a complementary module that can help enhance the ranking performance by structured traversal of the corpus graph. Hence, when the feedback signal from the ranker is strong, \gar{} is reproducible, offering performance gains even in a reasoning-intensive setup similar to the insights obtained in the original work \cite{macavaney2022adaptive} with marginal costs and no additional enhancements.

\subsubsection{ Ranking performance of \gar{} with weaker rankers}
For weaker rerankers (like MonoT5 and \llama{}), augmenting with \gar{} does not provide any gains or actually causes a deterioration in ranking performance across all evaluation sets in BRIGHT. Note that here, MonoT5 and \llama{} are weaker rerankers as their nDCG@10 is lower than the BM25 baseline performance as observed in Table \ref{tab:bright}. We posit that the drop in performance is primarily due to the poor relevance signal given by the reranker, which serves as feedback to traverse the corpus graph in \gar{}. Due to poor signal provided by these LLM rerankers, \gar{} prioritizes less relevant or noisy documents from the neighborhood instead of discovering more relevant documents, resulting in poor ranking performance. We observe a similar trend with \rank{}-0.5B, which also has a lower average nDCG@10 across the entire BRIGHT benchmark than the BM25 baseline. While, we observe that in the original \gar{} work MonoT5 based rankers lead to improved performance when augmented with \gar{} our results demonstrate that MonoT5 is not a strong ranker for reasoning intensive benchmark like BRIGHT. On analyzing results from the original paper, we observe that the approaches were primarily evaluated on TREC-DL test collections and MSMARCO passage ranking corpus which comprises on simpler web-based queries, and MonoT5 is an in-domain ranker trained on similar queries. These queries only require surface-level semantic matching, unlike the reasoning-intensive tasks in BRIGHT, which require the ranker to perform deeper reasoning, such as understanding the logic for code or theorem retrieval. Hence, MonoT5 variants perform poorly providing no gains or degradation of performance on such reasoning-intensive tasks, also demonstrating poor out-of-domain performance. This leads to low-quality feedback signals that prioritize noisy documents from the neighborhood graph instead of more relevant documents.

\definecolor{myColor1}{HTML}{1F77B4} %
\definecolor{myColor2}{HTML}{FF7F0E} %
\definecolor{myColor3}{HTML}{2CA02C} %
\definecolor{myColor4}{HTML}{9467BD} %

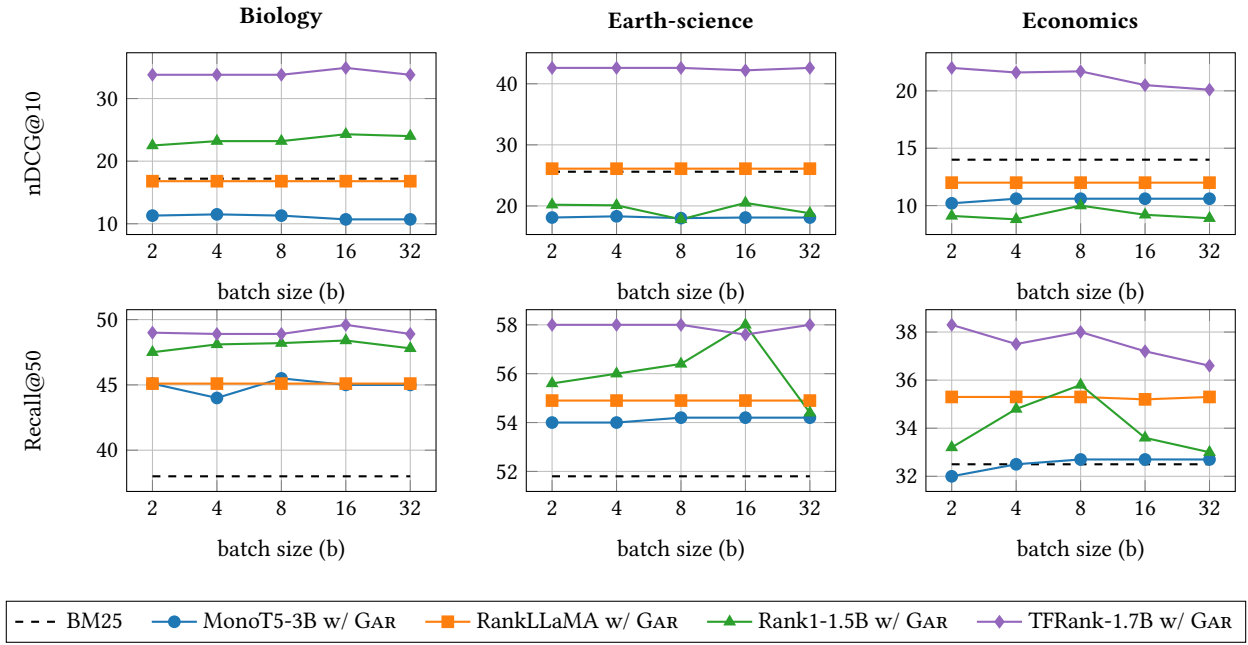
\begin{figure*}[t]
    \centering
    \begin{tikzpicture}
        \begin{groupplot}[
            group style={
                group size=3 by 2,      
                horizontal sep=1.2cm,   
                vertical sep=1.0cm,     
            },
            width=0.32\textwidth,
            height=4cm,
            xmode=log,                  
            log basis x=2,
            xtick={2,4,8,16,32},
            xticklabels={2,4,8,16,32},
            xlabel={batch size (b)},   
            grid=major,
            every axis plot/.append style={
                line width=0.5mm,
                mark size=2pt
            },
            cycle list={
                {black,dashed, line width = 0.8pt},
                {myColor1,mark=*, line width = 0.8pt},
                {myColor2,mark=square*, line width = 0.8pt},
                {myColor3,mark=triangle*, line width = 0.8pt},
                {myColor4,mark=diamond*, line width = 0.8pt}
            },
            legend columns=-1,
            legend entries={BM25, MonoT5-3B w/ \gar{}, RankLLaMA w/ \gar{}, Rank1-1.5B w/ \gar{}, TFRank-1.7B w/ \gar{}},
            legend to name=batch_named,
             legend style={
                 /tikz/every even column/.append style={column sep=0.3cm}
            }
        ]

        \nextgroupplot[ylabel={nDCG@10}, title={\textbf{Biology}}]
            \addplot coordinates {(2, 17.2) (4, 17.2) (8, 17.2) (16, 17.2) (32, 17.2)}; %
            \addplot coordinates {(2, 11.3) (4, 11.5) (8, 11.3) (16, 10.7) (32, 10.7)}; %
            \addplot coordinates {(2, 16.8) (4, 16.8) (8, 16.8) (16, 16.8) (32, 16.8)}; %
            \addplot coordinates {(2, 22.5) (4, 23.2) (8, 23.2) (16, 24.3) (32, 24.0)}; %
            \addplot coordinates {(2, 33.8) (4, 33.8) (8, 33.8) (16, 34.9) (32, 33.8)}; %

        \nextgroupplot[title={\textbf{Earth-science}}]
            \addplot coordinates {(2, 25.6) (4, 25.6) (8, 25.6) (16, 25.6) (32, 25.6)}; %
            \addplot coordinates {(2, 18.1) (4, 18.3) (8, 18.0) (16, 18.1) (32, 18.1)}; %
            \addplot coordinates {(2, 26.1) (4, 26.1) (8, 26.1) (16, 26.1) (32, 26.1)}; %
            \addplot coordinates {(2, 20.2) (4, 20.1) (8, 17.8) (16, 20.5) (32, 18.8)}; %
            \addplot coordinates {(2, 42.6) (4, 42.6) (8, 42.6) (16, 42.2) (32, 42.6)}; %

        \nextgroupplot[title={\textbf{Economics}}]
            \addplot coordinates {(2, 14) (4, 14) (8, 14) (16, 14) (32, 14)}; %
            \addplot coordinates {(2, 10.2) (4, 10.6) (8, 10.6) (16, 10.6) (32, 10.6)}; %
            \addplot coordinates {(2, 12.0) (4, 12.0) (8, 12.0) (16, 12.0) (32, 12.0)}; %
            \addplot coordinates {(2, 9.1) (4, 8.8) (8, 10.0) (16, 9.2) (32, 8.9)}; %
            \addplot coordinates {(2, 22.0) (4, 21.6) (8, 21.7) (16, 20.5) (32, 20.1)}; %
        \nextgroupplot[ylabel={Recall@50}]
            \addplot coordinates {(2, 38) (4, 38) (8, 38) (16, 38) (32, 38.0)}; %
            \addplot coordinates {(2, 45.1) (4, 44.0) (8, 45.5) (16, 45) (32, 45.0)}; %
            \addplot coordinates {(2, 45.1) (4, 45.1) (8, 45.1) (16, 45.1) (32, 45.1)}; %
            \addplot coordinates {(2, 47.5) (4, 48.1) (8, 48.2) (16, 48.4) (32, 47.8)}; %
            \addplot coordinates {(2, 49.0) (4, 48.9) (8, 48.9) (16, 49.6) (32, 48.9)}; %

        \nextgroupplot
            \addplot coordinates {(2, 51.8) (4, 51.8) (8, 51.8) (16, 51.8) (32, 51.8)}; %
            \addplot coordinates {(2, 54.0) (4, 54.0) (8, 54.2) (16, 54.2) (32, 54.2)}; %
            \addplot coordinates {(2, 54.9) (4, 54.9) (8, 54.9) (16, 54.9) (32, 54.9)}; %
            \addplot coordinates {(2, 55.6) (4, 56.0) (8, 56.4) (16, 58) (32, 54.4)}; %
            \addplot coordinates {(2, 58.0) (4, 58.0) (8, 58.0) (16, 57.6) (32, 58.0)}; %

        \nextgroupplot
            \addplot coordinates {(2, 32.5) (4, 32.5) (8, 32.5) (16, 32.5) (32, 32.5)}; %
            \addplot coordinates {(2, 32 ) (4, 32.5) (8, 32.7) (16, 32.7) (32, 32.7)}; %
            \addplot coordinates {(2, 35.3) (4, 35.3) (8, 35.3) (16, 35.2) (32, 35.3)}; %
            \addplot coordinates {(2, 33.2) (4, 34.8) (8, 35.8) (16, 33.6) (32, 33.0)}; %
            \addplot coordinates {(2, 38.3) (4, 37.5) (8, 38.0) (16, 37.2) (32, 36.6)}; %

        \end{groupplot}
    \end{tikzpicture}

\vspace{\baselineskip}
\begin{minipage}{\textwidth}
\begin{center}
\ref{batch_named}
\end{center}
\end{minipage} 
    
    \caption{Retrieval and ranking performance of \gar{} when augmented with different re-rankers and varying batch size $b$. The first row shows nDCG@10 and second shows Recall@50 performance.}
    \label{fig:batch_ablation}
\end{figure*}

\subsection{Ablation over Hyperparameters}
\label{sec:ablation}

To answer \textbf{RQ3}, we vary the different hyperparameters of \gar{} and report a detailed analysis of results. The pointwise rerankers can process the documents in batches, which makes them suitable for efficient reranking. Therefore, instead of reranking the batches sequentially from the initial retrieved results $R_0$, at each alternate iteration, it selects the batch from the graph frontier $F$. However, recent reasoning-based rankers, which are GPU-heavy, can not process a sufficiently large batch size. Therefore, it is important to evaluate the performance of \gar{} when the batch size varies from a small size to a bigger one. 

The batch size $b$ and number of neighbors $k$ in the corpus graph play a vital role in the \gar{} method. As the original paper empirically showed that \gar{} is robust to the selection of these hyperparameters. We reproduce \gar{} by varying batch size $b$ and neighbors $k$. In the first experiment, we fix the number of neighbors $k$ to $16$ and vary $b\in [2,4,8,16,32]$. We do not go beyond a batch size of $32$, since we have a ranking budget of $50$. We select a budget of $50$ due to limited resources, and also we observe that the \gar{} is effective at a lower budget, for example, $50$. We have a detailed discussion in Section~\ref{sec:reranking_budget}.

In the second experiment, we fix the batch size $b$ to $16$ and vary $k\in[2,4,8,16,32,64,128]$. Due to limited resources, we use MonoT5-3B, RankLLaMA-7B, \rank-1.5B, and TFRank-1.7B as rerankers. For evaluation, we use Biology, Earth-science, and Economics subsets of BRIGHT and report nDCG@10 and Recall@50 performance. We select only three subsets of BRIGHT due to limited resources and the expensive nature of reasoning rankers.

\subsubsection{Effect of Batch Size}
The batch size decides how many documents should be processed in an iteration, either from the initial ranking $R_0$ or from the graph frontier $F$. In Fig.~\ref{fig:batch_ablation}, we report the retrieval and ranking performance. We find that the w/ \gar{} pipeline remains stable with the selection of batch size $b$. The ranking performance (nDCG@10) remains in a ballpark over different batch sizes, which makes \gar{} highly effective even when resources are limited and can process smaller batches. Also, it makes it robust for 1-hop (when $b=32$) or multi-hop (when $b\leq16$) neighborhood traversal in the corpus graph. Though preparation and selection of batches might add some additional computational overhead, which is almost negligible compared to the computational cost of expensive rankers. On the other end, where the batch size is larger, for example $b=16$ or $32$, \gar{} selects only $18$ documents from the neighborhood (with at most 2-hop neighbors) and remains effective. For instance on Biology, at batch size $b=32$, Recall@50 is $48.9$ and nDCG@10 is $33.8$ with \tf-1.5B is 33.8, which are significant improvements over BM25 (nDCG@10 of 17.2) and the reranking baseline (nDCG@10 of 29.6). This is similar to the original observations made in the \gar{} work \cite{macavaney2022adaptive} where performance of \gar{} is found to be stable, when the batch size is sufficiently small ($b \le 128$).

\definecolor{myColor1}{HTML}{1F77B4} %
\definecolor{myColor2}{HTML}{FF7F0E} %
\definecolor{myColor3}{HTML}{2CA02C} %
\definecolor{myColor4}{HTML}{9467BD} %

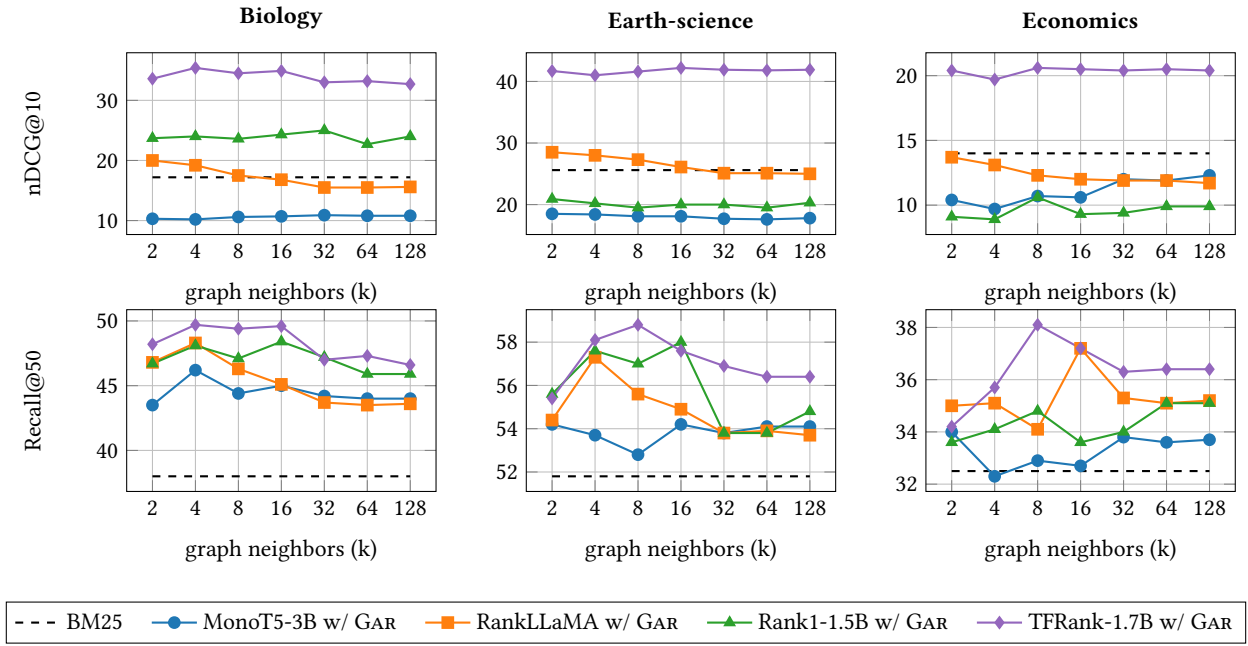
\begin{figure*}[t]
    \centering
    \begin{tikzpicture}
        \begin{groupplot}[
            group style={
                group size=3 by 2,      
                horizontal sep=1.2cm,   
                vertical sep=1.0cm
            },
            width=0.32\textwidth,
            height=4cm,
            xmode=log,                  
            log basis x=2,
            xtick={2,4,8,16,32,64, 128},
            xticklabels={2,4,8,16,32,64, 128},
            xlabel={graph neighbors (k)},   
            grid=major,
            every axis plot/.append style={
                line width=0.5mm,
                mark size=2pt
            },
            cycle list={
                {black,dashed, line width = 0.8pt},
                {myColor1,mark=*, line width = 0.8pt},
                {myColor2,mark=square*, line width = 0.8pt},
                {myColor3,mark=triangle*, line width = 0.8pt},
                {myColor4,mark=diamond*, line width = 0.8pt}
            },
            legend columns=-1,
            legend entries={BM25, MonoT5-3B w/ \gar{}, RankLLaMA w/ \gar{}, Rank1-1.5B w/ \gar{}, TFRank-1.7B w/ \gar{}},
            legend to name=lk_named,
             legend style={
                 /tikz/every even column/.append style={column sep=0.3cm}
            }
        ]

        \nextgroupplot[ylabel={nDCG@10}, title={\textbf{Biology}}]
            \addplot coordinates {(2, 17.2) (4, 17.2) (8, 17.2) (16, 17.2) (32, 17.2) (64, 17.2) (128, 17.2)}; %
            \addplot coordinates {(2, 10.3) (4, 10.2) (8, 10.6) (16, 10.7 ) (32, 10.9) (64, 10.8) (128, 10.8)}; %
            \addplot coordinates {(2, 20.0) (4, 19.2) (8, 17.5) (16, 16.8) (32, 15.5) (64, 15.5) (128, 15.6)}; %
            \addplot coordinates {(2, 23.7) (4, 24.0) (8, 23.6) (16, 24.3) (32, 25.0) (64, 22.7) (128, 24.0)}; %
            \addplot coordinates {(2, 33.6) (4, 35.4) (8, 34.5) (16, 34.9) (32, 33.0) (64, 33.2) (128, 32.7)}; %

        \nextgroupplot[title={\textbf{Earth-science}}]
            \addplot coordinates {(2, 25.6) (4, 25.6) (8, 25.6) (16, 25.6) (32, 25.6) (64, 25.6) (128, 25.6)}; %
            \addplot coordinates {(2, 18.5) (4, 18.4) (8, 18.1) (16, 18.1) (32, 17.7) (64, 17.6) (128, 17.8)}; %
            \addplot coordinates {(2, 28.5) (4, 28.0) (8, 27.3) (16, 26.1) (32, 25.1) (64, 25.1) (128, 25)}; %
            \addplot coordinates {(2, 20.9) (4, 20.2) (8, 19.5) (16, 20.0) (32, 20.0) (64, 19.5) (128, 20.3)}; %
            \addplot coordinates {(2, 41.7) (4, 41.0) (8, 41.6) (16, 42.2) (32, 41.9) (64, 41.8) (128,41.9)}; %

        \nextgroupplot[title={\textbf{Economics}}]
            \addplot coordinates {(2, 14) (4, 14) (8, 14) (16, 14) (32, 14) (64, 14) (128,14)}; %
            \addplot coordinates {(2, 10.4) (4, 9.7) (8, 10.7) (16, 10.6) (32, 12.0) (64, 11.9) (128, 12.3)}; %
            \addplot coordinates {(2, 13.7) (4, 13.1) (8, 12.3) (16, 12) (32, 11.9) (64, 11.9) (128, 11.7)}; %
            \addplot coordinates {(2, 9.1) (4, 8.9) (8, 10.6) (16, 9.3) (32, 9.4) (64, 9.9) (128, 9.9)}; %
            \addplot coordinates {(2, 20.4) (4, 19.7) (8, 20.6) (16, 20.5) (32, 20.4) (64, 20.5) (128, 20.4)}; %

        \nextgroupplot[ylabel={Recall@50}]
            \addplot coordinates {(2, 38) (4, 38) (8, 38) (16, 38) (32, 38.0) (64, 38.0) (128, 38.0) (128, 38)}; %

            \addplot coordinates {(2, 43.5) (4, 46.2) (8, 44.4) (16, 45) (32, 44.2) (64, 44.0) (128, 44.0)}; %
            \addplot coordinates {(2, 46.8) (4, 48.3) (8, 46.3) (16, 45.1) (32, 43.7) (64, 43.5) (128, 43.6)}; %
            \addplot coordinates {(2, 46.7) (4, 48.1) (8, 47.1) (16, 48.4) (32, 47.2) (64, 45.9) (128, 45.9)}; %
            \addplot coordinates {(2, 48.2) (4, 49.7) (8, 49.4) (16, 49.6) (32, 47.0) (64, 47.3) (128, 46.6)}; %

        \nextgroupplot
            \addplot coordinates {(2, 51.8) (4, 51.8) (8, 51.8) (16, 51.8) (32, 51.8) (64, 51.8) (128, 51.8)}; %

            \addplot coordinates {(2, 54.2) (4, 53.7) (8, 52.8) (16, 54.2) (32, 53.8) (64, 54.1) (128, 54.1)}; %
            \addplot coordinates {(2, 54.4) (4, 57.3) (8, 55.6) (16, 54.9) (32, 53.8) (64, 53.9) (128, 53.7)}; %
            \addplot coordinates {(2, 55.6) (4, 57.6) (8, 57.0) (16, 58.0) (32, 53.8) (64, 53.8) (128, 54.8)}; %
            \addplot coordinates {(2, 55.4) (4, 58.1) (8, 58.8) (16, 57.6) (32, 56.9) (64, 56.4) (128, 56.4)}; %

        \nextgroupplot
            \addplot coordinates {(2, 32.5) (4, 32.5) (8, 32.5) (16, 32.5) (32, 32.5) (64, 32.5) (128, 32.5)}; %
            \addplot coordinates {(2, 34.0) (4, 32.3) (8, 32.9) (16, 32.7) (32, 33.8) (64, 33.6) (128, 33.7)}; %
            \addplot coordinates {(2, 35.0) (4, 35.1) (8, 34.1) (16, 37.2) (32, 35.3) (64, 35.1) (128, 35.2)}; %
            \addplot coordinates {(2, 33.6) (4, 34.1) (8, 34.8) (16, 33.6) (32, 34.0) (64, 35.1) (128, 35.1)}; %
            \addplot coordinates {(2, 34.2) (4, 35.7) (8, 38.1) (16, 37.2) (32, 36.3) (64, 36.4) (128, 36.4)}; %

        \end{groupplot}
    \end{tikzpicture}

\vspace{\baselineskip}
\begin{minipage}{\textwidth}
\begin{center}
\ref{lk_named}
\end{center}
\end{minipage} 
    
    \caption{Effect of number of neighbors in the corpus graph. The first row shows nDCG@10 performance, and the second row Recall@50.}
    \label{fig:lk_ablation}
\end{figure*}

\subsubsection{Effect of Neighborhood Size}
In the search system, we hope that if we explore more, the chances of finding further relevant documents are higher. Standard reranking methods focus on exploring documents in the retrieval depth; however, \gar{} also focus on exploring the documents that are close (neighbors in the corpus graph) to initially retrieved documents. The number of neighbors decides how many documents from a source document should be added in graph frontier $F$. For instance, at $k=16$, for each reranked document, 16 neighbors are added to $F$ and get scores of their corresponding source document. The original paper did ablation for $1\leq k \leq 16$. We also investigate some queries manually and observe relevant documents appear in the deeper ($k>80$) neighborhood, also shown in Figure~\ref{fig:query_analysis}. Therefore, we choose higher values of $k$ too.

We report the performance of \gar{} by varying the number of neighbors in Figure~\ref{fig:lk_ablation}. Most of the rerankers show a drop in Recall@50 when the number of neighbors increases, especially in the range of $16$ to $128$. A major drawback of the \gar{} method is that it can not differentiate between neighbors, since all neighbors get the same scores in the graph frontier. This is also observed in previous work~\cite{rathee2024quam}. In theory, we should find more relevant documents, however \gar{} can not filter out noisy neighbors. Further, these neural rankers are not \textit{oracle} rankers, they can also rank an irrelevant document higher and then \gar{} end up adding all noisy neighbors in the graph frontier. On Biology, the Recall peaks in the range of $4$ to $16$. The most degradation happens for \llama, and its side effect can be seen in nDCG@10 performance. We also report average nDCG@10 on BRIGHT by \llama{} in Figure~\ref{fig:ndcg_c50}. We observe that the feedback signals from \llama{} do not help in finding relevant documents in the neighborhood. Further, we observe that \gar{} is most robust with  \tf-1.7B, which is a good reranker for reasoning-intensive tasks. Therefore, even if it provides high-quality relevance scores, the design of \gar{} algorithm forces to choose documents that are deeper into the neighborhood of only the top 1-2 documents. The neighbors of other documents might not get a chance to be selected. Even after selecting these noisy documents (as shown by the Recall@50 performance), it still ranks them lower in the final ranked list, and therefore, nDCG@10 also remains stable. 

Surprisingly, the performance of \mono-3b either remains stable or improves further as we increase the number of neighbors. We posit this is due to its overall performance at reranking budget $50$, where nDCG@10 improves (contrary to budget 100). 

Overall, our experimental findings align with the original paper; the \gar{} performance remains stable at higher graph depth and peaks at lower depth.

\subsection{Reranking Budget}
\label{sec:reranking_budget}
\begin{figure}[h!]
    \centering
    \begin{tikzpicture}
        \begin{axis}[
            ybar,
            bar width=5pt,
            width=1.1\columnwidth,
            height=6cm, 
            title={Avg. nDCG@10 on BRIGHT},
            ylabel near ticks,
            symbolic x coords={MonoT5-base, MonoT5-3B, RankLLaMA-7B, Qwen-0.6B, Qwen-4B, TFRank-0.6B, TFRank-1.7B, TFRank-4B, TFRank-8B, Rank-0.5B, Rank-1.5B, Rank1-7B},
            xtick=data,
            xticklabel style={rotate=45, anchor=east, align=right, font=\footnotesize},
            ymin=0,
            ymax=28,
    legend style={at={(0.25,0.98)}, anchor=north east, nodes={scale=0.8, transform shape}, font=\small },
    nodes near coords,
    every node near coord/.append style={font=\tiny, /pgf/number format/fixed, /pgf/number format/precision=1},
    legend cell align={left},
    axis line style={black},
    major tick style={black},
    tick label style={font=\small},
    label style={font=\small},
    enlarge x limits=0.05,
        ]

        \addplot[fill=blue!40] coordinates {
            (MonoT5-base,8.99) 
            (MonoT5-3B,9.49) 
            (RankLLaMA-7B,13.93) 
            (Qwen-0.6B,14.90) 
            (Qwen-4B,21.15) 
            (TFRank-0.6B,16.48) 
            (TFRank-1.7B,19.04) 
            (TFRank-4B,21.74) 
            (TFRank-8B,21.85) 
            (Rank-0.5B,9.44) 
            (Rank-1.5B,12.07)
            (Rank1-7B, 18.24)
        };
        \addlegendentry{Base Model}

        \addplot[fill=orange!40] coordinates {
            (MonoT5-base,9.22) 
            (MonoT5-3B,9.80) 
            (RankLLaMA-7B,13.27) 
            (Qwen-0.6B,15.96) 
            (Qwen-4B,22.89) 
            (TFRank-0.6B,18.75) 
            (TFRank-1.7B,21.18) 
            (TFRank-4B,24.00) 
            (TFRank-8B,24.40) 
            (Rank-0.5B,9.12) 
            (Rank-1.5B,12.35)
            (Rank1-7B, 19.63)
        };
        \addlegendentry{w/ \gar{}}

        \draw [black, dashed, thick] 
            ({rel axis cs:0,0}|-{axis cs:MonoT5-base,12.6}) -- 
            ({rel axis cs:1,0}|-{axis cs:MonoT5-base,12.6}) 
            node [pos=0.02, above right, black, font=\tiny] {BM25 (12.6)};

        \end{axis}
    \end{tikzpicture}
    \caption{Average nDCG@10 comparison of various ranking models on BRIGHT when top-$50$ retrieved documents from BM25 are reranked. The blue bars represent the base model performance, while the orange bars show the performance with GAR augmentation. The dashed line indicates the BM25 baseline score of 12.6.}
    \label{fig:ndcg_c50}
\end{figure}
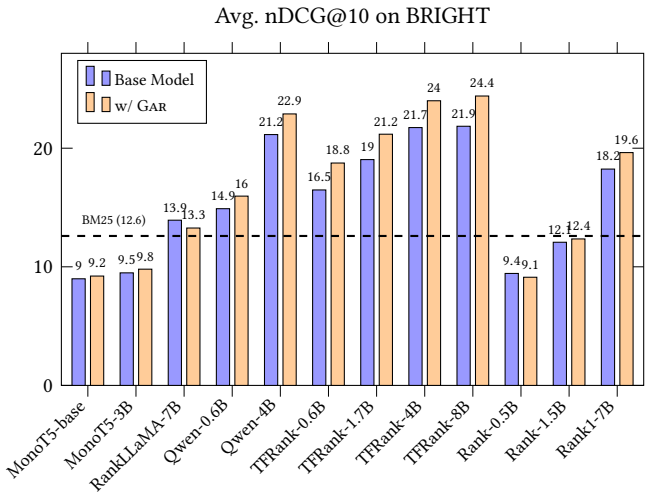

The computational cost and power-hungry behavior~\cite{greenir} of these expensive (both reasoning and non-reasoning) rankers make it practically impossible to process a larger list of documents. Also, in the extreme case, scoring more documents exhaustively results in performance degradation due to hard negatives~\cite{jacob2024drowning}. So, it is highly beneficial if we can have high effectiveness at lower reranking budgets. Since \gar{} goes beyond sequential reranking by exploring the corpus graph and using the same ranking costs, we reproduce \gar{} on BRIGHT when we have a small reranking budget, for instance, $c=50$. We report\footnote{We report subtask-level performance in the code repository.} average nDCG@10 by different rerankers in Figure~\ref{fig:ndcg_c50}. We observe that \gar{} remain highly effective at a lower budget as well. For instance, \tf-0.6B with \gar{} improves average nDCG@10 from 16.5 to 18.8 and \tf-4B from 21.7 to 24. 

More interestingly, we observe that the performance of the reranking pipeline, which reranks $100$ documents, can be achieved by reranking just $50$ documents (0.5x), which reduces the computational cost, latency, and carbon footprint. For instance, on Biology, nDCG@10 by \tf-1.7B after reranking $100$ documents is 32.3. On the other hand, \tf-1.7B with \gar{} achieves nDCG@10 of $34.9$. 

Another interesting observation we make is the performance of weaker rerankers such as \mono-base, 3B, \llama-7B, \rank-0.5B, or \rank-1.5B. At a higher budget, where \gar{} chooses more documents from the corpus graph, due to the weak signals from these rerankers, \gar{} selects irrelevant documents and therefore performance degrades (as shown in Section~\ref{sec:effectiveness}), however, at a low ranking budget of $50$, \gar{} selects comparatively fewer documents from the corpus graph and \gar{} does not degrade performance.  

\subsection{Models Comparison}
\label{sec:model_comparison}
\begin{figure}
    \centering
    \begin{tikzpicture}
        \begin{axis}[
            width=8cm,
            height=5.2cm,
            xlabel={Model Size},
            ylabel={nDCG@10},
            grid=major,
            grid style={dashed, gray!30},
            symbolic x coords={0.5B, 0.6B, 1.5B, 1.7B, 3B, 4B, 7B, 8B},
            xtick={0.5B, 0.6B, 1.5B, 1.7B, 3B, 4B, 7B, 8B},
            xticklabel style={font=\large},
            ymin=5, ymax=30,
            legend style={
                at={(0.5,-0.6)}, 
                anchor=south,
                legend columns=4,
                draw=none,
                font=\small, 
                row sep=0.1cm,
                /tikz/every even column/.append style={column sep=0.0cm}
            },
        ]
        \addplot[color=black, dashed, thick, forget plot] 
            coordinates {(0.5B, 12.58) (0.6B, 12.58) (1.5B, 12.58) (1.7B, 12.58) (3B, 12.58) (4B, 12.58) (7B, 12.58) (8B, 12.58)} node[pos=0.8, above] {BM25}; 
        
        \addplot[color=blue!60!black, mark=*, dashed, thick] 
            coordinates {(0.6B, 14.7) (4B, 20.82) (8B, 20.82)};
        \addlegendentry{Qwen}
        
        \addplot[color=blue!60!black, mark=*, thick, mark options={fill=white}] 
            coordinates {(0.6B, 15.30) (4B, 22.79) (8B, 22.8)};
        \addlegendentry{Qwen w/ \gar{}}

        \addplot[color=teal, mark=square*, dashed, thick] 
            coordinates {(0.6B, 16.45) (1.7B, 19.68) (4B, 23.02) (8B, 23.67)};
        \addlegendentry{TFRank}

        \addplot[color=teal, mark=square*, thick, mark options={fill=white}] 
            coordinates {(0.6B, 17.97) (1.7B, 21.72) (4B, 25.39) (8B, 26.52)};
        \addlegendentry{TFRank w/ \gar{}}

        \addplot[color=orange, mark=triangle*, dashed, thick] 
            coordinates {(0.5B, 7.65) (1.5B, 11.09) (7B, 19.68)};
        \addlegendentry{\rank{}}

        \addplot[color=orange, mark=triangle*, thick, mark options={fill=white}] 
            coordinates {(0.5B, 6.83) (1.5B, 10.37) (7B, 20.98)};
        \addlegendentry{\rank{} w/ \gar{}}

        \end{axis}
    \end{tikzpicture}
    \caption{Performance comparison (avg. nDCG@10) on BRIGHT of reasoning models of different scales and their corresponding GAR variants.}
    \Description{A line chart showing performance metrics for four models}
    \label{fig:models_scale}
    
\end{figure}
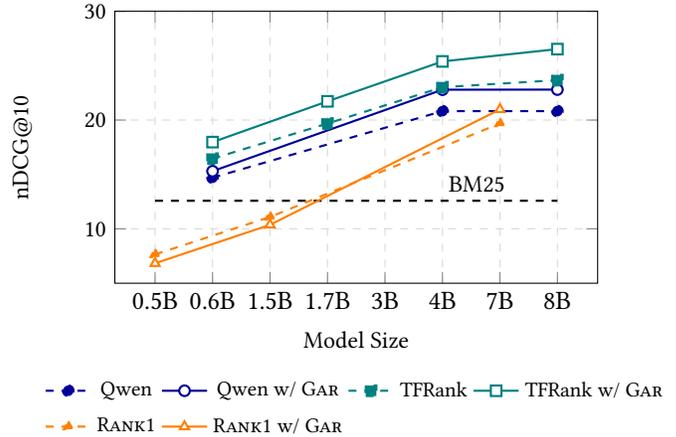

To answer \textbf{RQ2} partially, we study the behaviour of \gar{} with reasoning rankers, \qwen{}, \tf{}, and \rank{} of different scales, in terms of the number of parameters. The size of the reasoning model plays a crucial role in processing the query-document pairs, especially in terms of latency and the hardware needed to run the experiments. The goal of this experiment is to see how \gar{} performance varies with the size of the ranking model. The key hypothesis is that performance should scale with the size of the ranking model. We report the average nDCG@10 by rerankers of different sizes in Figure~\ref{fig:models_scale} when the reranking budget is $100$, and the original query is used for retrieval.

\definecolor{acmblue}{RGB}{0, 114, 178}
\definecolor{acmorange}{RGB}{213, 94, 0}
\definecolor{acmgreen}{RGB}{0, 158, 115}
\definecolor{acmpurple}{RGB}{204, 121, 167}

The first observation we make is that \gar{} does not show performance gains with the small scale ranker, \rank-0.5B and \rank-1.5B, but rather, it shows a drop in performance. We again posit that this is due to their poor performance in the ranking task, since the average nDCG@10 is worse than the BM25 baseline as discussed in Section \ref{sec:results_rq1}. On the other hand, the better ranking models of similar size from \qwen{} and \tf{}, mainly \qwen-0.6B, \tf-0.6B, and \tf-1.7B, have the better ranking performance than BM25. \gar{} shows improvements when used with such rerankers. For example, \gar{} improves the performance of \tf-1.7b from $19.7$ to $21.7$. However, the gains for \qwen-0.6B are comparatively lower. Further, at 4B size, both \qwen-4B and \tf-4B show performance gains when \gar{} is used. Further, the 7B variant of \rank \ demonstrates performance gains when augmented with \gar{}, as it shows stronger ranking capabilities over BM25. \gar{} improves nDCG@10 from $19.7$ to $21$. Moreover, where the performance of \qwen{} saturates as we go from 4B to 8B variant, the gains from \gar{} also saturate. Contrary to this, \tf{} rankers still improve nDCG@10 from scale of 4B to 8B, and hence corresponding performance when augmented with \gar{} also improves. At 8B, \tf{} with \gar{} improves nDCG@10 from 23.7 to 26.5. 

Another observation we make is that the ranking performance of a larger scale model can be achieved by using a smaller scale model when used with \gar{}. For instance, \tf-1.7B with \gar{} achieves better performance than \qwen-4B, 8B, and \rank-7B. This demonstrates the effectiveness of \gar{} in resource-constrained scenarios, such as when one is limited to running a 1.7B parameter model. 

Another benefit of \gar{} is its minimal computational overhead. As shown in the original paper, \gar{} adds only 3–4ms of latency per 100 documents, negligible compared to the latency of reasoning rankers. We do not perform a separate latency costs study in this work; since these overheads are independent of the specific task or ranker, the cost is driven solely by graph frontier updates and alternative batch selection. Indeed, the total cost is dominated by the reranker (especially for large reasoning and thinking tasks), and we observed no discernible difference in runtime between systems with and without \gar{}.

\section{Conclusion}
We reproduce \gar{}, Graph-based Adaptive Re-ranking for reasoning-intensive information retrieval benchmark, BRIGHT. We consider a variety of reasoning and non-reasoning rankers. We find that \gar{} helps in improving both retrieval and ranking performance on reasoning-intensive IR task. It serves as a training-free module that is complementary to other enhancements such as using a large scale ranker. However, there is strong correlation between the performance of base ranker and the performance of \gar{}. If the feedback signals from ranker are not weak, \gar{} can negatively impact the performance. Performance of \gar{} also depends on the other type of feedback that comes from the quality of initial retrieval stage and type of the corpus graph. We leave it for future study. Further, we observe that \gar{} is robust to the selection of hyperparameters, especially when used with strong rankers. Finally, we observe that a small scale strong ranker with \gar{} can achieve the ranking performance of the large scale ranker.

\section*{Acknowledgments}
This work was partially funded by the Bundesministerium für Wirtschaft und Energie (BMWE), Germany, in the context of the 8ra Initiative ("Soofi", 13IPC040E), and by the European Union’s Horizon Europe Research and Innovation Programme JustREACH under Grant Agreement No 101214666. 

\bibliographystyle{ACM-Reference-Format}
\balance
\bibliography{bib}


\begin{thebibliography}{49}

%%% the Web address.

\ifx \showCODEN    \undefined \def \showCODEN     #1{\unskip}     \fi
\ifx \showISBNx    \undefined \def \showISBNx     #1{\unskip}     \fi
\ifx \showISBNxiii \undefined \def \showISBNxiii  #1{\unskip}     \fi
\ifx \showISSN     \undefined \def \showISSN      #1{\unskip}     \fi
\ifx \showLCCN     \undefined \def \showLCCN      #1{\unskip}     \fi
\ifx \shownote     \undefined \def \shownote      #1{#1}          \fi
\ifx \showarticletitle \undefined \def \showarticletitle #1{#1}   \fi
\ifx \showURL      \undefined \def \showURL       {\relax}        \fi
% invisible to TeX
\providecommand\bibfield[2]{#2}
\providecommand\bibinfo[2]{#2}
\providecommand\natexlab[1]{#1}
\providecommand\showeprint[2][]{arXiv:#2}

\bibitem[An et~al\mbox{.}(2026)]%
        {an2026fastinsight}
\bibfield{author}{\bibinfo{person}{Seonho An}, \bibinfo{person}{Chaejeong Hyun}, {and} \bibinfo{person}{Min-Soo Kim}.} \bibinfo{year}{2026}\natexlab{}.
\newblock \showarticletitle{FastInsight: Fast and Insightful Retrieval via Fusion Operators for Graph RAG}.
\newblock \bibinfo{journal}{\emph{arXiv preprint arXiv:2601.18579}} (\bibinfo{year}{2026}).
\newblock


\bibitem[Asai et~al\mbox{.}(2023)]%
        {asai-etal-2023-task}
\bibfield{author}{\bibinfo{person}{Akari Asai}, \bibinfo{person}{Timo Schick}, \bibinfo{person}{Patrick Lewis}, \bibinfo{person}{Xilun Chen}, \bibinfo{person}{Gautier Izacard}, \bibinfo{person}{Sebastian Riedel}, \bibinfo{person}{Hannaneh Hajishirzi}, {and} \bibinfo{person}{Wen-tau Yih}.} \bibinfo{year}{2023}\natexlab{}.
\newblock \showarticletitle{Task-aware Retrieval with Instructions}. In \bibinfo{booktitle}{\emph{Findings of the Association for Computational Linguistics: ACL 2023}}, \bibfield{editor}{\bibinfo{person}{Anna Rogers}, \bibinfo{person}{Jordan Boyd-Graber}, {and} \bibinfo{person}{Naoaki Okazaki}} (Eds.). \bibinfo{publisher}{Association for Computational Linguistics}, \bibinfo{address}{Toronto, Canada}, \bibinfo{pages}{3650--3675}.
\newblock
\href{https://doi.org/10.18653/v1/2023.findings-acl.225}{doi:\nolinkurl{10.18653/v1/2023.findings-acl.225}}


\bibitem[Bruch et~al\mbox{.}(2023)]%
        {bruch2023analysis}
\bibfield{author}{\bibinfo{person}{Sebastian Bruch}, \bibinfo{person}{Siyu Gai}, {and} \bibinfo{person}{Amir Ingber}.} \bibinfo{year}{2023}\natexlab{}.
\newblock \showarticletitle{An analysis of fusion functions for hybrid retrieval}.
\newblock \bibinfo{journal}{\emph{ACM Transactions on Information Systems}} \bibinfo{volume}{42}, \bibinfo{number}{1} (\bibinfo{year}{2023}), \bibinfo{pages}{1--35}.
\newblock


\bibitem[Chen et~al\mbox{.}(2022)]%
        {chen2022out}
\bibfield{author}{\bibinfo{person}{Tao Chen}, \bibinfo{person}{Mingyang Zhang}, \bibinfo{person}{Jing Lu}, \bibinfo{person}{Michael Bendersky}, {and} \bibinfo{person}{Marc Najork}.} \bibinfo{year}{2022}\natexlab{}.
\newblock \showarticletitle{Out-of-domain semantics to the rescue! zero-shot hybrid retrieval models}. In \bibinfo{booktitle}{\emph{European Conference on Information Retrieval}}. Springer, \bibinfo{pages}{95--110}.
\newblock


\bibitem[Chen et~al\mbox{.}(2025)]%
        {chenbrowsecomp}
\bibfield{author}{\bibinfo{person}{Zijian Chen}, \bibinfo{person}{Xueguang Ma}, \bibinfo{person}{Shengyao Zhuang}, \bibinfo{person}{Ping Nie}, \bibinfo{person}{Kai Zou}, \bibinfo{person}{Sahel Sharifymoghaddam}, \bibinfo{person}{Andrew Liu}, \bibinfo{person}{Joshua Green}, \bibinfo{person}{Kshama Patel}, \bibinfo{person}{Ruoxi Meng}, {et~al\mbox{.}}} \bibinfo{year}{2025}\natexlab{}.
\newblock \showarticletitle{BrowseComp-Plus: A More Fair and Transparent Evaluation Benchmark of Deep-Research Agent}. In \bibinfo{booktitle}{\emph{First Workshop on Multi-Turn Interactions in Large Language Models}}.
\newblock


\bibitem[Cormack et~al\mbox{.}(2009)]%
        {cormack2009reciprocal}
\bibfield{author}{\bibinfo{person}{Gordon~V. Cormack}, \bibinfo{person}{Charles L~A Clarke}, {and} \bibinfo{person}{Stefan Buettcher}.} \bibinfo{year}{2009}\natexlab{}.
\newblock \showarticletitle{Reciprocal rank fusion outperforms condorcet and individual rank learning methods}. In \bibinfo{booktitle}{\emph{Proceedings of the 32nd International ACM SIGIR Conference on Research and Development in Information Retrieval}} (Boston, MA, USA) \emph{(\bibinfo{series}{SIGIR '09})}. \bibinfo{publisher}{Association for Computing Machinery}, \bibinfo{address}{New York, NY, USA}, \bibinfo{pages}{758–759}.
\newblock
\showISBNx{9781605584836}
\href{https://doi.org/10.1145/1571941.1572114}{doi:\nolinkurl{10.1145/1571941.1572114}}


\bibitem[Dunn et~al\mbox{.}(2025)]%
        {approximate2025dunn}
\bibfield{author}{\bibinfo{person}{Lachlan Dunn}, \bibinfo{person}{Luke Gallagher}, {and} \bibinfo{person}{Joel Mackenzie}.} \bibinfo{year}{2025}\natexlab{}.
\newblock \showarticletitle{Approximate Bag-of-Words Top-k Corpus Graphs}. In \bibinfo{booktitle}{\emph{Advances in Information Retrieval}}, \bibfield{editor}{\bibinfo{person}{Claudia Hauff}, \bibinfo{person}{Craig Macdonald}, \bibinfo{person}{Dietmar Jannach}, \bibinfo{person}{Gabriella Kazai}, \bibinfo{person}{Franco~Maria Nardini}, \bibinfo{person}{Fabio Pinelli}, \bibinfo{person}{Fabrizio Silvestri}, {and} \bibinfo{person}{Nicola Tonellotto}} (Eds.). \bibinfo{publisher}{Springer Nature Switzerland}, \bibinfo{address}{Cham}, \bibinfo{pages}{174--182}.
\newblock
\showISBNx{978-3-031-88714-7}


\bibitem[Edge et~al\mbox{.}(2024)]%
        {edge2024local}
\bibfield{author}{\bibinfo{person}{Darren Edge}, \bibinfo{person}{Ha Trinh}, \bibinfo{person}{Newman Cheng}, \bibinfo{person}{Joshua Bradley}, \bibinfo{person}{Alex Chao}, \bibinfo{person}{Apurva Mody}, \bibinfo{person}{Steven Truitt}, \bibinfo{person}{Dasha Metropolitansky}, \bibinfo{person}{Robert~Osazuwa Ness}, {and} \bibinfo{person}{Jonathan Larson}.} \bibinfo{year}{2024}\natexlab{}.
\newblock \showarticletitle{From local to global: A graph rag approach to query-focused summarization}.
\newblock \bibinfo{journal}{\emph{arXiv preprint arXiv:2404.16130}} (\bibinfo{year}{2024}).
\newblock


\bibitem[Fan et~al\mbox{.}(2025)]%
        {fan2025tfrankthinkfreereasoningenables}
\bibfield{author}{\bibinfo{person}{Yongqi Fan}, \bibinfo{person}{Xiaoyang Chen}, \bibinfo{person}{Dezhi Ye}, \bibinfo{person}{Jie Liu}, \bibinfo{person}{Haijin Liang}, \bibinfo{person}{Jin Ma}, \bibinfo{person}{Ben He}, \bibinfo{person}{Yingfei Sun}, {and} \bibinfo{person}{Tong Ruan}.} \bibinfo{year}{2025}\natexlab{}.
\newblock \bibinfo{title}{TFRank: Think-Free Reasoning Enables Practical Pointwise LLM Ranking}.
\newblock
\showeprint[arxiv]{2508.09539}~[cs.IR]
\urldef\tempurl%
\url{https://arxiv.org/abs/2508.09539}
\showURL{%
\tempurl}


\bibitem[Formal et~al\mbox{.}(2021)]%
        {splade}
\bibfield{author}{\bibinfo{person}{Thibault Formal}, \bibinfo{person}{Benjamin Piwowarski}, {and} \bibinfo{person}{St\'{e}phane Clinchant}.} \bibinfo{year}{2021}\natexlab{}.
\newblock \showarticletitle{SPLADE: Sparse Lexical and Expansion Model for First Stage Ranking}. In \bibinfo{booktitle}{\emph{Proceedings of the 44th International ACM SIGIR Conference on Research and Development in Information Retrieval}} (<conf-loc>, <city>Virtual Event</city>, <country>Canada</country>, </conf-loc>) \emph{(\bibinfo{series}{SIGIR '21})}. \bibinfo{publisher}{Association for Computing Machinery}, \bibinfo{address}{New York, NY, USA}, \bibinfo{pages}{2288–2292}.
\newblock
\showISBNx{9781450380379}
\href{https://doi.org/10.1145/3404835.3463098}{doi:\nolinkurl{10.1145/3404835.3463098}}


\bibitem[Guo et~al\mbox{.}(2025)]%
        {guo2025deepseek}
\bibfield{author}{\bibinfo{person}{Daya Guo}, \bibinfo{person}{Dejian Yang}, \bibinfo{person}{Haowei Zhang}, \bibinfo{person}{Junxiao Song}, \bibinfo{person}{Ruoyu Zhang}, \bibinfo{person}{Runxin Xu}, \bibinfo{person}{Qihao Zhu}, \bibinfo{person}{Shirong Ma}, \bibinfo{person}{Peiyi Wang}, \bibinfo{person}{Xiao Bi}, {et~al\mbox{.}}} \bibinfo{year}{2025}\natexlab{}.
\newblock \showarticletitle{Deepseek-r1: Incentivizing reasoning capability in llms via reinforcement learning}.
\newblock \bibinfo{journal}{\emph{arXiv preprint arXiv:2501.12948}} (\bibinfo{year}{2025}).
\newblock


\bibitem[Jacob et~al\mbox{.}(2024)]%
        {jacob2024drowning}
\bibfield{author}{\bibinfo{person}{Mathew Jacob}, \bibinfo{person}{Erik Lindgren}, \bibinfo{person}{Matei Zaharia}, \bibinfo{person}{Michael Carbin}, \bibinfo{person}{Omar Khattab}, {and} \bibinfo{person}{Andrew Drozdov}.} \bibinfo{year}{2024}\natexlab{}.
\newblock \showarticletitle{Drowning in documents: consequences of scaling reranker inference}.
\newblock \bibinfo{journal}{\emph{arXiv preprint arXiv:2411.11767}} (\bibinfo{year}{2024}).
\newblock


\bibitem[Jardine and van Rijsbergen(1971)]%
        {jardine1971use}
\bibfield{author}{\bibinfo{person}{N. Jardine} {and} \bibinfo{person}{Cornelis~Joost van Rijsbergen}.} \bibinfo{year}{1971}\natexlab{}.
\newblock \showarticletitle{The use of hierarchic clustering in information retrieval}.
\newblock \bibinfo{journal}{\emph{Inf. Storage Retr.}} \bibinfo{volume}{7}, \bibinfo{number}{5} (\bibinfo{year}{1971}), \bibinfo{pages}{217--240}.
\newblock
\href{https://doi.org/10.1016/0020-0271(71)90051-9}{doi:\nolinkurl{10.1016/0020-0271(71)90051-9}}


\bibitem[Karpukhin et~al\mbox{.}(2020)]%
        {karpukhin-etal-2020-dense}
\bibfield{author}{\bibinfo{person}{Vladimir Karpukhin}, \bibinfo{person}{Barlas Oguz}, \bibinfo{person}{Sewon Min}, \bibinfo{person}{Patrick Lewis}, \bibinfo{person}{Ledell Wu}, \bibinfo{person}{Sergey Edunov}, \bibinfo{person}{Danqi Chen}, {and} \bibinfo{person}{Wen-tau Yih}.} \bibinfo{year}{2020}\natexlab{}.
\newblock \showarticletitle{Dense Passage Retrieval for Open-Domain Question Answering}. In \bibinfo{booktitle}{\emph{Proceedings of the 2020 Conference on Empirical Methods in Natural Language Processing (EMNLP)}}, \bibfield{editor}{\bibinfo{person}{Bonnie Webber}, \bibinfo{person}{Trevor Cohn}, \bibinfo{person}{Yulan He}, {and} \bibinfo{person}{Yang Liu}} (Eds.). \bibinfo{publisher}{Association for Computational Linguistics}, \bibinfo{address}{Online}, \bibinfo{pages}{6769--6781}.
\newblock
\href{https://doi.org/10.18653/v1/2020.emnlp-main.550}{doi:\nolinkurl{10.18653/v1/2020.emnlp-main.550}}


\bibitem[Killingback and Zamani(2025)]%
        {killingback2025benchmarking}
\bibfield{author}{\bibinfo{person}{Julian Killingback} {and} \bibinfo{person}{Hamed Zamani}.} \bibinfo{year}{2025}\natexlab{}.
\newblock \showarticletitle{Benchmarking Information Retrieval Models on Complex Retrieval Tasks}.
\newblock \bibinfo{journal}{\emph{arXiv preprint arXiv:2509.07253}} (\bibinfo{year}{2025}).
\newblock


\bibitem[Kim et~al\mbox{.}(2026)]%
        {kim2026adaptive}
\bibfield{author}{\bibinfo{person}{Jongho Kim}, \bibinfo{person}{Jaeyoung Kim}, \bibinfo{person}{Seung-won Hwang}, \bibinfo{person}{Jihyuk Kim}, \bibinfo{person}{Yu~Jin Kim}, {and} \bibinfo{person}{Moontae Lee}.} \bibinfo{year}{2026}\natexlab{}.
\newblock \showarticletitle{Adaptive Retrieval for Reasoning-Intensive Retrieval}.
\newblock \bibinfo{journal}{\emph{arXiv preprint arXiv:2601.04618}} (\bibinfo{year}{2026}).
\newblock


\bibitem[Kulkarni et~al\mbox{.}(2024)]%
        {Kulkarni2024lexboost}
\bibfield{author}{\bibinfo{person}{Hrishikesh Kulkarni}, \bibinfo{person}{Nazli Goharian}, \bibinfo{person}{Ophir Frieder}, {and} \bibinfo{person}{Sean MacAvaney}.} \bibinfo{year}{2024}\natexlab{}.
\newblock \showarticletitle{LexBoost: Improving Lexical Document Retrieval with Nearest Neighbors}. In \bibinfo{booktitle}{\emph{Proceedings of the ACM Symposium on Document Engineering 2024}} (San Jose, CA, USA) \emph{(\bibinfo{series}{DocEng '24})}. \bibinfo{publisher}{Association for Computing Machinery}, \bibinfo{address}{New York, NY, USA}, Article \bibinfo{articleno}{16}, \bibinfo{numpages}{10}~pages.
\newblock
\showISBNx{9798400711695}
\href{https://doi.org/10.1145/3685650.3685658}{doi:\nolinkurl{10.1145/3685650.3685658}}


\bibitem[Kulkarni et~al\mbox{.}(2023)]%
        {kulkarni2023lexically}
\bibfield{author}{\bibinfo{person}{Hrishikesh Kulkarni}, \bibinfo{person}{Sean MacAvaney}, \bibinfo{person}{Nazli Goharian}, {and} \bibinfo{person}{Ophir Frieder}.} \bibinfo{year}{2023}\natexlab{}.
\newblock \showarticletitle{Lexically-Accelerated Dense Retrieval}. In \bibinfo{booktitle}{\emph{Proceedings of the 46th International {ACM} {SIGIR} Conference on Research and Development in Information Retrieval, {SIGIR} 2023, Taipei, Taiwan, July 23-27, 2023}}, \bibfield{editor}{\bibinfo{person}{Hsin{-}Hsi Chen}, \bibinfo{person}{Wei{-}Jou~(Edward) Duh}, \bibinfo{person}{Hen{-}Hsen Huang}, \bibinfo{person}{Makoto~P. Kato}, \bibinfo{person}{Josiane Mothe}, {and} \bibinfo{person}{Barbara Poblete}} (Eds.). \bibinfo{publisher}{{ACM}}, \bibinfo{pages}{152--162}.
\newblock
\href{https://doi.org/10.1145/3539618.3591715}{doi:\nolinkurl{10.1145/3539618.3591715}}


\bibitem[Lan et~al\mbox{.}(2026)]%
        {lan2026retro}
\bibfield{author}{\bibinfo{person}{Junwei Lan}, \bibinfo{person}{Jianlyu Chen}, \bibinfo{person}{Zheng Liu}, \bibinfo{person}{Chaofan Li}, \bibinfo{person}{Siqi Bao}, {and} \bibinfo{person}{Defu Lian}.} \bibinfo{year}{2026}\natexlab{}.
\newblock \showarticletitle{Retro*: Optimizing {LLM}s for Reasoning-Intensive Document Retrieval}. In \bibinfo{booktitle}{\emph{The Fourteenth International Conference on Learning Representations}}.
\newblock
\urldef\tempurl%
\url{https://openreview.net/forum?id=0WGl8PNMSA}
\showURL{%
\tempurl}


\bibitem[Lee et~al\mbox{.}(2025)]%
        {lee2025imagine}
\bibfield{author}{\bibinfo{person}{Sangam Lee}, \bibinfo{person}{Ryang Heo}, \bibinfo{person}{SeongKu Kang}, {and} \bibinfo{person}{Dongha Lee}.} \bibinfo{year}{2025}\natexlab{}.
\newblock \showarticletitle{Imagine All The Relevance: Scenario-Profiled Indexing with Knowledge Expansion for Dense Retrieval}.
\newblock \bibinfo{journal}{\emph{arXiv preprint arXiv:2503.23033}} (\bibinfo{year}{2025}).
\newblock


\bibitem[Lin et~al\mbox{.}(2021)]%
        {lin2021batch}
\bibfield{author}{\bibinfo{person}{Sheng{-}Chieh Lin}, \bibinfo{person}{Jheng{-}Hong Yang}, {and} \bibinfo{person}{Jimmy Lin}.} \bibinfo{year}{2021}\natexlab{}.
\newblock \showarticletitle{In-Batch Negatives for Knowledge Distillation with Tightly-Coupled Teachers for Dense Retrieval}. In \bibinfo{booktitle}{\emph{Proceedings of the 6th Workshop on Representation Learning for NLP, RepL4NLP@ACL-IJCNLP 2021, Online, August 6, 2021}}, \bibfield{editor}{\bibinfo{person}{Anna Rogers}, \bibinfo{person}{Iacer Calixto}, \bibinfo{person}{Ivan Vulic}, \bibinfo{person}{Naomi Saphra}, \bibinfo{person}{Nora Kassner}, \bibinfo{person}{Oana{-}Maria Camburu}, \bibinfo{person}{Trapit Bansal}, {and} \bibinfo{person}{Vered Shwartz}} (Eds.). \bibinfo{publisher}{Association for Computational Linguistics}, \bibinfo{pages}{163--173}.
\newblock
\href{https://doi.org/10.18653/V1/2021.REPL4NLP-1.17}{doi:\nolinkurl{10.18653/V1/2021.REPL4NLP-1.17}}


\bibitem[Ma et~al\mbox{.}(2024)]%
        {ma2024fine}
\bibfield{author}{\bibinfo{person}{Xueguang Ma}, \bibinfo{person}{Liang Wang}, \bibinfo{person}{Nan Yang}, \bibinfo{person}{Furu Wei}, {and} \bibinfo{person}{Jimmy Lin}.} \bibinfo{year}{2024}\natexlab{}.
\newblock \showarticletitle{Fine-tuning llama for multi-stage text retrieval}. In \bibinfo{booktitle}{\emph{Proceedings of the 47th International ACM SIGIR Conference on Research and Development in Information Retrieval}}. \bibinfo{pages}{2421--2425}.
\newblock


\bibitem[MacAvaney et~al\mbox{.}(2020)]%
        {DBLP:conf/sigir/MacAvaneyN0TGF20b}
\bibfield{author}{\bibinfo{person}{Sean MacAvaney}, \bibinfo{person}{Franco~Maria Nardini}, \bibinfo{person}{Raffaele Perego}, \bibinfo{person}{Nicola Tonellotto}, \bibinfo{person}{Nazli Goharian}, {and} \bibinfo{person}{Ophir Frieder}.} \bibinfo{year}{2020}\natexlab{}.
\newblock \showarticletitle{Expansion via Prediction of Importance with Contextualization}. In \bibinfo{booktitle}{\emph{Proceedings of the 43rd International {ACM} {SIGIR} conference on research and development in Information Retrieval, {SIGIR} 2020, Virtual Event, China, July 25-30, 2020}}, \bibfield{editor}{\bibinfo{person}{Jimmy~X. Huang}, \bibinfo{person}{Yi~Chang}, \bibinfo{person}{Xueqi Cheng}, \bibinfo{person}{Jaap Kamps}, \bibinfo{person}{Vanessa Murdock}, \bibinfo{person}{Ji{-}Rong Wen}, {and} \bibinfo{person}{Yiqun Liu}} (Eds.). \bibinfo{publisher}{{ACM}}, \bibinfo{pages}{1573--1576}.
\newblock
\href{https://doi.org/10.1145/3397271.3401262}{doi:\nolinkurl{10.1145/3397271.3401262}}


\bibitem[MacAvaney et~al\mbox{.}(2022)]%
        {macavaney2022adaptive}
\bibfield{author}{\bibinfo{person}{Sean MacAvaney}, \bibinfo{person}{Nicola Tonellotto}, {and} \bibinfo{person}{Craig Macdonald}.} \bibinfo{year}{2022}\natexlab{}.
\newblock \showarticletitle{Adaptive Re-Ranking with a Corpus Graph}. In \bibinfo{booktitle}{\emph{Proceedings of the 31st {ACM} International Conference on Information {\&} Knowledge Management, Atlanta, GA, USA, October 17-21, 2022}}, \bibfield{editor}{\bibinfo{person}{Mohammad~Al Hasan} {and} \bibinfo{person}{Li~Xiong}} (Eds.). \bibinfo{publisher}{{ACM}}, \bibinfo{pages}{1491--1500}.
\newblock
\href{https://doi.org/10.1145/3511808.3557231}{doi:\nolinkurl{10.1145/3511808.3557231}}


\bibitem[Mallia et~al\mbox{.}(2019)]%
        {DBLP:conf/sigir/MalliaSMS19}
\bibfield{author}{\bibinfo{person}{Antonio Mallia}, \bibinfo{person}{Michal Siedlaczek}, \bibinfo{person}{Joel~M. Mackenzie}, {and} \bibinfo{person}{Torsten Suel}.} \bibinfo{year}{2019}\natexlab{}.
\newblock \showarticletitle{{PISA:} Performant Indexes and Search for Academia}. In \bibinfo{booktitle}{\emph{Proceedings of the Open-Source {IR} Replicability Challenge co-located with 42nd International {ACM} {SIGIR} Conference on Research and Development in Information Retrieval, OSIRRC@SIGIR 2019, Paris, France, July 25, 2019}} \emph{(\bibinfo{series}{{CEUR} Workshop Proceedings}, Vol.~\bibinfo{volume}{2409})}, \bibfield{editor}{\bibinfo{person}{Ryan Clancy}, \bibinfo{person}{Nicola Ferro}, \bibinfo{person}{Claudia Hauff}, \bibinfo{person}{Jimmy Lin}, \bibinfo{person}{Tetsuya Sakai}, {and} \bibinfo{person}{Ze~Zhong Wu}} (Eds.). \bibinfo{publisher}{CEUR-WS.org}, \bibinfo{pages}{50--56}.
\newblock
\urldef\tempurl%
\url{https://ceur-ws.org/Vol-2409/docker08.pdf}
\showURL{%
\tempurl}


\bibitem[Muennighoff et~al\mbox{.}(2025)]%
        {muennighoff2025generative}
\bibfield{author}{\bibinfo{person}{Niklas Muennighoff}, \bibinfo{person}{Hongjin SU}, \bibinfo{person}{Liang Wang}, \bibinfo{person}{Nan Yang}, \bibinfo{person}{Furu Wei}, \bibinfo{person}{Tao Yu}, \bibinfo{person}{Amanpreet Singh}, {and} \bibinfo{person}{Douwe Kiela}.} \bibinfo{year}{2025}\natexlab{}.
\newblock \showarticletitle{Generative Representational Instruction Tuning}. In \bibinfo{booktitle}{\emph{The Thirteenth International Conference on Learning Representations}}.
\newblock
\urldef\tempurl%
\url{https://openreview.net/forum?id=BC4lIvfSzv}
\showURL{%
\tempurl}


\bibitem[Nogueira et~al\mbox{.}(2020)]%
        {nogueira2020document}
\bibfield{author}{\bibinfo{person}{Rodrigo~Frassetto Nogueira}, \bibinfo{person}{Zhiying Jiang}, \bibinfo{person}{Ronak Pradeep}, {and} \bibinfo{person}{Jimmy Lin}.} \bibinfo{year}{2020}\natexlab{}.
\newblock \showarticletitle{Document Ranking with a Pretrained Sequence-to-Sequence Model}. In \bibinfo{booktitle}{\emph{Findings of the Association for Computational Linguistics: {EMNLP} 2020, Online Event, 16-20 November 2020}} \emph{(\bibinfo{series}{Findings of {ACL}}, Vol.~\bibinfo{volume}{{EMNLP} 2020})}, \bibfield{editor}{\bibinfo{person}{Trevor Cohn}, \bibinfo{person}{Yulan He}, {and} \bibinfo{person}{Yang Liu}} (Eds.). \bibinfo{publisher}{Association for Computational Linguistics}, \bibinfo{pages}{708--718}.
\newblock
\href{https://doi.org/10.18653/V1/2020.FINDINGS-EMNLP.63}{doi:\nolinkurl{10.18653/V1/2020.FINDINGS-EMNLP.63}}


\bibitem[Pradeep et~al\mbox{.}(2023)]%
        {pradeep2023rankzephyr}
\bibfield{author}{\bibinfo{person}{Ronak Pradeep}, \bibinfo{person}{Sahel Sharifymoghaddam}, {and} \bibinfo{person}{Jimmy Lin}.} \bibinfo{year}{2023}\natexlab{}.
\newblock \showarticletitle{RankZephyr: Effective and Robust Zero-Shot Listwise Reranking is a Breeze!}
\newblock \bibinfo{journal}{\emph{arXiv preprint arXiv:2312.02724}} (\bibinfo{year}{2023}).
\newblock


\bibitem[Rathee et~al\mbox{.}(2025a)]%
        {guiding2025rathee}
\bibfield{author}{\bibinfo{person}{Mandeep Rathee}, \bibinfo{person}{Sean MacAvaney}, {and} \bibinfo{person}{Avishek Anand}.} \bibinfo{year}{2025}\natexlab{a}.
\newblock \showarticletitle{Guiding Retrieval Using LLM-Based Listwise Rankers}. In \bibinfo{booktitle}{\emph{Advances in Information Retrieval}}, \bibfield{editor}{\bibinfo{person}{Claudia Hauff}, \bibinfo{person}{Craig Macdonald}, \bibinfo{person}{Dietmar Jannach}, \bibinfo{person}{Gabriella Kazai}, \bibinfo{person}{Franco~Maria Nardini}, \bibinfo{person}{Fabio Pinelli}, \bibinfo{person}{Fabrizio Silvestri}, {and} \bibinfo{person}{Nicola Tonellotto}} (Eds.). \bibinfo{publisher}{Springer Nature Switzerland}, \bibinfo{address}{Cham}, \bibinfo{pages}{230--246}.
\newblock
\showISBNx{978-3-031-88708-6}


\bibitem[Rathee et~al\mbox{.}(2025b)]%
        {rathee2024quam}
\bibfield{author}{\bibinfo{person}{Mandeep Rathee}, \bibinfo{person}{Sean MacAvaney}, {and} \bibinfo{person}{Avishek Anand}.} \bibinfo{year}{2025}\natexlab{b}.
\newblock \showarticletitle{Quam: Adaptive Retrieval through Query Affinity Modelling}. In \bibinfo{booktitle}{\emph{Proceedings of the Eighteenth ACM International Conference on Web Search and Data Mining}} (Hannover, Germany) \emph{(\bibinfo{series}{WSDM '25})}. \bibinfo{publisher}{Association for Computing Machinery}, \bibinfo{address}{New York, NY, USA}, \bibinfo{pages}{954–962}.
\newblock
\showISBNx{9798400713293}
\href{https://doi.org/10.1145/3701551.3703584}{doi:\nolinkurl{10.1145/3701551.3703584}}


\bibitem[Rathee et~al\mbox{.}(2025c)]%
        {ore}
\bibfield{author}{\bibinfo{person}{Mandeep Rathee}, \bibinfo{person}{Venktesh V}, \bibinfo{person}{Sean MacAvaney}, {and} \bibinfo{person}{Avishek Anand}.} \bibinfo{year}{2025}\natexlab{c}.
\newblock \showarticletitle{Breaking the Lens of the Telescope: Online Relevance Estimation over Large Retrieval Sets}. In \bibinfo{booktitle}{\emph{Proceedings of the 48th International ACM SIGIR Conference on Research and Development in Information Retrieval}} (Padua, Italy) \emph{(\bibinfo{series}{SIGIR '25})}. \bibinfo{publisher}{Association for Computing Machinery}, \bibinfo{address}{New York, NY, USA}, \bibinfo{pages}{2287–2297}.
\newblock
\showISBNx{9798400715921}
\href{https://doi.org/10.1145/3726302.3729910}{doi:\nolinkurl{10.1145/3726302.3729910}}


\bibitem[Robertson(2008)]%
        {DBLP:journals/jis/Robertson08}
\bibfield{author}{\bibinfo{person}{Stephen Robertson}.} \bibinfo{year}{2008}\natexlab{}.
\newblock \showarticletitle{On the history of evaluation in {IR}}.
\newblock \bibinfo{journal}{\emph{J. Inf. Sci.}} \bibinfo{volume}{34}, \bibinfo{number}{4} (\bibinfo{year}{2008}), \bibinfo{pages}{439--456}.
\newblock
\href{https://doi.org/10.1177/0165551507086989}{doi:\nolinkurl{10.1177/0165551507086989}}


\bibitem[Robertson and Zaragoza(2009)]%
        {bm25}
\bibfield{author}{\bibinfo{person}{Stephen Robertson} {and} \bibinfo{person}{Hugo Zaragoza}.} \bibinfo{year}{2009}\natexlab{}.
\newblock \showarticletitle{The Probabilistic Relevance Framework: BM25 and Beyond}.
\newblock \bibinfo{journal}{\emph{Found. Trends Inf. Retr.}} \bibinfo{volume}{3}, \bibinfo{number}{4} (\bibinfo{date}{apr} \bibinfo{year}{2009}), \bibinfo{pages}{333–389}.
\newblock
\showISSN{1554-0669}
\href{https://doi.org/10.1561/1500000019}{doi:\nolinkurl{10.1561/1500000019}}


\bibitem[Scells et~al\mbox{.}(2022)]%
        {greenir}
\bibfield{author}{\bibinfo{person}{Harrisen Scells}, \bibinfo{person}{Shengyao Zhuang}, {and} \bibinfo{person}{Guido Zuccon}.} \bibinfo{year}{2022}\natexlab{}.
\newblock \showarticletitle{Reduce, Reuse, Recycle: Green Information Retrieval Research}. In \bibinfo{booktitle}{\emph{Proceedings of the 45th International ACM SIGIR Conference on Research and Development in Information Retrieval}} (Madrid, Spain) \emph{(\bibinfo{series}{SIGIR '22})}. \bibinfo{publisher}{Association for Computing Machinery}, \bibinfo{address}{New York, NY, USA}, \bibinfo{pages}{2825–2837}.
\newblock
\showISBNx{9781450387323}
\href{https://doi.org/10.1145/3477495.3531766}{doi:\nolinkurl{10.1145/3477495.3531766}}


\bibitem[Sinhababu et~al\mbox{.}(2024)]%
        {sinhababu-etal-2024-shot}
\bibfield{author}{\bibinfo{person}{Nilanjan Sinhababu}, \bibinfo{person}{Andrew Parry}, \bibinfo{person}{Debasis Ganguly}, \bibinfo{person}{Debasis Samanta}, {and} \bibinfo{person}{Pabitra Mitra}.} \bibinfo{year}{2024}\natexlab{}.
\newblock \showarticletitle{Few-shot Prompting for Pairwise Ranking: An Effective Non-Parametric Retrieval Model}. In \bibinfo{booktitle}{\emph{Findings of the Association for Computational Linguistics: EMNLP 2024}}, \bibfield{editor}{\bibinfo{person}{Yaser Al-Onaizan}, \bibinfo{person}{Mohit Bansal}, {and} \bibinfo{person}{Yun-Nung Chen}} (Eds.). \bibinfo{publisher}{Association for Computational Linguistics}, \bibinfo{address}{Miami, Florida, USA}, \bibinfo{pages}{12363--12377}.
\newblock
\href{https://doi.org/10.18653/v1/2024.findings-emnlp.720}{doi:\nolinkurl{10.18653/v1/2024.findings-emnlp.720}}


\bibitem[SU et~al\mbox{.}(2025)]%
        {su2025bright}
\bibfield{author}{\bibinfo{person}{Hongjin SU}, \bibinfo{person}{Howard Yen}, \bibinfo{person}{Mengzhou Xia}, \bibinfo{person}{Weijia Shi}, \bibinfo{person}{Niklas Muennighoff}, \bibinfo{person}{Han yu Wang}, \bibinfo{person}{Liu Haisu}, \bibinfo{person}{Quan Shi}, \bibinfo{person}{Zachary~S Siegel}, \bibinfo{person}{Michael Tang}, \bibinfo{person}{Ruoxi Sun}, \bibinfo{person}{Jinsung Yoon}, \bibinfo{person}{Sercan~O Arik}, \bibinfo{person}{Danqi Chen}, {and} \bibinfo{person}{Tao Yu}.} \bibinfo{year}{2025}\natexlab{}.
\newblock \showarticletitle{{BRIGHT}: A Realistic and Challenging Benchmark for Reasoning-Intensive Retrieval}. In \bibinfo{booktitle}{\emph{The Thirteenth International Conference on Learning Representations}}.
\newblock
\urldef\tempurl%
\url{https://openreview.net/forum?id=ykuc5q381b}
\showURL{%
\tempurl}


\bibitem[Sun et~al\mbox{.}(2023)]%
        {sun2023chatgpt}
\bibfield{author}{\bibinfo{person}{Weiwei Sun}, \bibinfo{person}{Lingyong Yan}, \bibinfo{person}{Xinyu Ma}, \bibinfo{person}{Shuaiqiang Wang}, \bibinfo{person}{Pengjie Ren}, \bibinfo{person}{Zhumin Chen}, \bibinfo{person}{Dawei Yin}, {and} \bibinfo{person}{Zhaochun Ren}.} \bibinfo{year}{2023}\natexlab{}.
\newblock \showarticletitle{Is ChatGPT good at search? investigating large language models as re-ranking agents}.
\newblock \bibinfo{journal}{\emph{arXiv preprint arXiv:2304.09542}} (\bibinfo{year}{2023}).
\newblock


\bibitem[Thakur et~al\mbox{.}(2025)]%
        {thakur2025freshstack}
\bibfield{author}{\bibinfo{person}{Nandan Thakur}, \bibinfo{person}{Jimmy Lin}, \bibinfo{person}{Sam Havens}, \bibinfo{person}{Michael Carbin}, \bibinfo{person}{Omar Khattab}, {and} \bibinfo{person}{Andrew Drozdov}.} \bibinfo{year}{2025}\natexlab{}.
\newblock \showarticletitle{FreshStack: Building Realistic Benchmarks for Evaluating Retrieval on Technical Documents}. In \bibinfo{booktitle}{\emph{The Thirty-ninth Annual Conference on Neural Information Processing Systems Datasets and Benchmarks Track}}.
\newblock
\urldef\tempurl%
\url{https://openreview.net/forum?id=54TTgXlS2U}
\showURL{%
\tempurl}


\bibitem[V et~al\mbox{.}(2025)]%
        {v-etal-2025-sunar}
\bibfield{author}{\bibinfo{person}{Venktesh V}, \bibinfo{person}{Mandeep Rathee}, {and} \bibinfo{person}{Avishek Anand}.} \bibinfo{year}{2025}\natexlab{}.
\newblock \showarticletitle{{SUNAR}: Semantic Uncertainty based Neighborhood Aware Retrieval for Complex {QA}}. In \bibinfo{booktitle}{\emph{Proceedings of the 2025 Conference of the Nations of the Americas Chapter of the Association for Computational Linguistics: Human Language Technologies (Volume 1: Long Papers)}}, \bibfield{editor}{\bibinfo{person}{Luis Chiruzzo}, \bibinfo{person}{Alan Ritter}, {and} \bibinfo{person}{Lu~Wang}} (Eds.). \bibinfo{publisher}{Association for Computational Linguistics}, \bibinfo{address}{Albuquerque, New Mexico}, \bibinfo{pages}{5818--5835}.
\newblock
\showISBNx{979-8-89176-189-6}
\href{https://doi.org/10.18653/v1/2025.naacl-long.300}{doi:\nolinkurl{10.18653/v1/2025.naacl-long.300}}


\bibitem[Wang et~al\mbox{.}(2011)]%
        {wang2011cascade}
\bibfield{author}{\bibinfo{person}{Lidan Wang}, \bibinfo{person}{Jimmy Lin}, {and} \bibinfo{person}{Donald Metzler}.} \bibinfo{year}{2011}\natexlab{}.
\newblock \showarticletitle{A cascade ranking model for efficient ranked retrieval}. In \bibinfo{booktitle}{\emph{Proceedings of the 34th international ACM SIGIR conference on Research and development in Information Retrieval}}. \bibinfo{pages}{105--114}.
\newblock


\bibitem[Wang et~al\mbox{.}(2021)]%
        {wang2021bert}
\bibfield{author}{\bibinfo{person}{Shuai Wang}, \bibinfo{person}{Shengyao Zhuang}, {and} \bibinfo{person}{Guido Zuccon}.} \bibinfo{year}{2021}\natexlab{}.
\newblock \showarticletitle{Bert-based dense retrievers require interpolation with bm25 for effective passage retrieval}. In \bibinfo{booktitle}{\emph{Proceedings of the 2021 ACM SIGIR international conference on theory of information retrieval}}. \bibinfo{pages}{317--324}.
\newblock


\bibitem[Wei et~al\mbox{.}(2022)]%
        {wei2022chain}
\bibfield{author}{\bibinfo{person}{Jason Wei}, \bibinfo{person}{Xuezhi Wang}, \bibinfo{person}{Dale Schuurmans}, \bibinfo{person}{Maarten Bosma}, \bibinfo{person}{brian ichter}, \bibinfo{person}{Fei Xia}, \bibinfo{person}{Ed Chi}, \bibinfo{person}{Quoc~V Le}, {and} \bibinfo{person}{Denny Zhou}.} \bibinfo{year}{2022}\natexlab{}.
\newblock \showarticletitle{Chain-of-Thought Prompting Elicits Reasoning in Large Language Models}. In \bibinfo{booktitle}{\emph{Advances in Neural Information Processing Systems}}, \bibfield{editor}{\bibinfo{person}{S.~Koyejo}, \bibinfo{person}{S.~Mohamed}, \bibinfo{person}{A.~Agarwal}, \bibinfo{person}{D.~Belgrave}, \bibinfo{person}{K.~Cho}, {and} \bibinfo{person}{A.~Oh}} (Eds.), Vol.~\bibinfo{volume}{35}. \bibinfo{publisher}{Curran Associates, Inc.}, \bibinfo{pages}{24824--24837}.
\newblock
\urldef\tempurl%
\url{https://proceedings.neurips.cc/paper_files/paper/2022/file/9d5609613524ecf4f15af0f7b31abca4-Paper-Conference.pdf}
\showURL{%
\tempurl}


\bibitem[Weller et~al\mbox{.}(2025a)]%
        {weller-etal-2025-followir}
\bibfield{author}{\bibinfo{person}{Orion Weller}, \bibinfo{person}{Benjamin Chang}, \bibinfo{person}{Sean MacAvaney}, \bibinfo{person}{Kyle Lo}, \bibinfo{person}{Arman Cohan}, \bibinfo{person}{Benjamin Van~Durme}, \bibinfo{person}{Dawn Lawrie}, {and} \bibinfo{person}{Luca Soldaini}.} \bibinfo{year}{2025}\natexlab{a}.
\newblock \showarticletitle{{F}ollow{IR}: Evaluating and Teaching Information Retrieval Models to Follow Instructions}. In \bibinfo{booktitle}{\emph{Proceedings of the 2025 Conference of the Nations of the Americas Chapter of the Association for Computational Linguistics: Human Language Technologies (Volume 1: Long Papers)}}, \bibfield{editor}{\bibinfo{person}{Luis Chiruzzo}, \bibinfo{person}{Alan Ritter}, {and} \bibinfo{person}{Lu~Wang}} (Eds.). \bibinfo{publisher}{Association for Computational Linguistics}, \bibinfo{address}{Albuquerque, New Mexico}, \bibinfo{pages}{11926--11942}.
\newblock
\showISBNx{979-8-89176-189-6}
\href{https://doi.org/10.18653/v1/2025.naacl-long.597}{doi:\nolinkurl{10.18653/v1/2025.naacl-long.597}}


\bibitem[Weller et~al\mbox{.}(2025b)]%
        {DBLP:conf/iclr/WellerDLPZH25}
\bibfield{author}{\bibinfo{person}{Orion Weller}, \bibinfo{person}{Benjamin~Van Durme}, \bibinfo{person}{Dawn~J. Lawrie}, \bibinfo{person}{Ashwin Paranjape}, \bibinfo{person}{Yuhao Zhang}, {and} \bibinfo{person}{Jack Hessel}.} \bibinfo{year}{2025}\natexlab{b}.
\newblock \showarticletitle{Promptriever: Instruction-Trained Retrievers Can Be Prompted Like Language Models}. In \bibinfo{booktitle}{\emph{The Thirteenth International Conference on Learning Representations, {ICLR} 2025, Singapore, April 24-28, 2025}}. \bibinfo{publisher}{OpenReview.net}.
\newblock
\urldef\tempurl%
\url{https://openreview.net/forum?id=odvSjn416y}
\showURL{%
\tempurl}


\bibitem[Weller et~al\mbox{.}(2025c)]%
        {weller2025rank1}
\bibfield{author}{\bibinfo{person}{Orion Weller}, \bibinfo{person}{Kathryn Ricci}, \bibinfo{person}{Eugene Yang}, \bibinfo{person}{Andrew Yates}, \bibinfo{person}{Dawn Lawrie}, {and} \bibinfo{person}{Benjamin Van~Durme}.} \bibinfo{year}{2025}\natexlab{c}.
\newblock \showarticletitle{Rank1: Test-time compute for reranking in information retrieval}.
\newblock \bibinfo{journal}{\emph{arXiv preprint arXiv:2502.18418}} (\bibinfo{year}{2025}).
\newblock


\bibitem[Yoon et~al\mbox{.}(2025)]%
        {yoon2025listwise}
\bibfield{author}{\bibinfo{person}{Soyoung Yoon}, \bibinfo{person}{Jongho Kim}, \bibinfo{person}{Daeyong Kwon}, \bibinfo{person}{Avishek Anand}, {and} \bibinfo{person}{Seung-won Hwang}.} \bibinfo{year}{2025}\natexlab{}.
\newblock \showarticletitle{On Listwise Reranking for Corpus Feedback}.
\newblock \bibinfo{journal}{\emph{arXiv preprint arXiv:2510.00887}} (\bibinfo{year}{2025}).
\newblock


\bibitem[Zhang et~al\mbox{.}(2025b)]%
        {zhang-etal-2025-rearank}
\bibfield{author}{\bibinfo{person}{Le Zhang}, \bibinfo{person}{Bo Wang}, \bibinfo{person}{Xipeng Qiu}, \bibinfo{person}{Siva Reddy}, {and} \bibinfo{person}{Aishwarya Agrawal}.} \bibinfo{year}{2025}\natexlab{b}.
\newblock \showarticletitle{{REARANK}: Reasoning Re-ranking Agent via Reinforcement Learning}. In \bibinfo{booktitle}{\emph{Proceedings of the 2025 Conference on Empirical Methods in Natural Language Processing}}, \bibfield{editor}{\bibinfo{person}{Christos Christodoulopoulos}, \bibinfo{person}{Tanmoy Chakraborty}, \bibinfo{person}{Carolyn Rose}, {and} \bibinfo{person}{Violet Peng}} (Eds.). \bibinfo{publisher}{Association for Computational Linguistics}, \bibinfo{address}{Suzhou, China}, \bibinfo{pages}{2458--2471}.
\newblock
\showISBNx{979-8-89176-332-6}
\href{https://doi.org/10.18653/v1/2025.emnlp-main.125}{doi:\nolinkurl{10.18653/v1/2025.emnlp-main.125}}


\bibitem[Zhang et~al\mbox{.}(2025a)]%
        {qwen3embedding}
\bibfield{author}{\bibinfo{person}{Yanzhao Zhang}, \bibinfo{person}{Mingxin Li}, \bibinfo{person}{Dingkun Long}, \bibinfo{person}{Xin Zhang}, \bibinfo{person}{Huan Lin}, \bibinfo{person}{Baosong Yang}, \bibinfo{person}{Pengjun Xie}, \bibinfo{person}{An Yang}, \bibinfo{person}{Dayiheng Liu}, \bibinfo{person}{Junyang Lin}, \bibinfo{person}{Fei Huang}, {and} \bibinfo{person}{Jingren Zhou}.} \bibinfo{year}{2025}\natexlab{a}.
\newblock \showarticletitle{Qwen3 Embedding: Advancing Text Embedding and Reranking Through Foundation Models}.
\newblock \bibinfo{journal}{\emph{arXiv preprint arXiv:2506.05176}} (\bibinfo{year}{2025}).
\newblock


\bibitem[Zhuang et~al\mbox{.}(2025)]%
        {rankr1}
\bibfield{author}{\bibinfo{person}{Shengyao Zhuang}, \bibinfo{person}{Xueguang Ma}, \bibinfo{person}{Bevan Koopman}, \bibinfo{person}{Jimmy Lin}, {and} \bibinfo{person}{Guido Zuccon}.} \bibinfo{year}{2025}\natexlab{}.
\newblock \bibinfo{title}{Rank-R1: Enhancing Reasoning in LLM-based Document Rerankers via Reinforcement Learning}.
\newblock
\showeprint[arxiv]{2503.06034}~[cs.IR]
\urldef\tempurl%
\url{https://arxiv.org/abs/2503.06034}
\showURL{%
\tempurl}


\end{thebibliography}

\end{document}